\documentclass[
  aps,prx,reprint,superscriptaddress,
  longbibliography 
]{revtex4-2}

\usepackage{graphicx}
\usepackage{amsmath,amssymb,bm,mathtools}
\usepackage{xcolor}
\usepackage{tikz}
\usetikzlibrary{arrows.meta,positioning,calc}
\usepackage{float}
\usepackage{hyperref}
\hypersetup{colorlinks=true,linkcolor=blue,citecolor=blue,urlcolor=blue}

\begin{document}

\title{Graph-Recurrent Learning of Emergent Collective Dynamics}

\author{Ali Rayat}
\affiliation{Department of Physics, University of Virginia, Charlottesville, Virginia 22904, USA}

\author{Acacia Lee}
\affiliation{Department of Physics, University of Virginia, Charlottesville, Virginia 22904, USA}

\author{Gia-Wei Chern}
\affiliation{Department of Physics, University of Virginia, Charlottesville, Virginia 22904, USA}

\date{\today}

\begin{abstract}
Collective ordering in correlated electron systems emerges from microscopic dynamics, yet the evolution of observable textures need not admit a closed, memoryless description. We introduce a graph-recurrent neural network (GRNN) framework that learns such emergent dynamics directly from trajectories without specifying microscopic equations of motion. The framework integrates gated recurrent memory with graph message passing: local hidden states encode information beyond the instantaneous observables, while spatial coupling allows this information to propagate between sites. The observable and hidden fields evolve jointly through a shared local rule, providing a scalable representation of dynamics with spatial interactions and temporal memory. We demonstrate this framework on the emergent dynamics of charge-density-wave order in the semiclassical Holstein model, successfully reproducing temporal and spatial correlations and dynamical scaling during domain coarsening. By learning finite-time evolution directly, the GRNN offers a route to efficient, data-driven descriptions of collective dynamics with unresolved microscopic degrees of freedom.
\end{abstract}

\maketitle

Pattern formation is a pervasive feature of correlated electron materials, where competing interactions generate magnetic textures, phase-separated clusters, and electronic stripes~\cite{moreo1999,moreo2000,white1998,zheng2017,huang2017}. The evolution of these patterns poses a further challenge. Domains grow and defects move and annihilate across widely separated length and time scales, often exhibiting dynamical scaling during phase ordering~\cite{bray1994,hohenberg1977}. At the same time, their motion can depend on the electronic degrees of freedom underlying the competing orders. Phenomenological descriptions, including time-dependent Ginzburg--Landau and phase-field models, reach large scales efficiently but may omit these material-specific effects~\cite{allen1979,cahn1958,chen2002}. Microscopic simulations retain the electronic degrees of freedom, but must recalculate their response as the configuration evolves, restricting accessible sizes and times~\cite{barros2013,chern2018}. Machine learning (ML) offers a route to efficient, transferable dynamical models that connect microscopic physics with large-scale pattern formation.

A prominent example is the development of ML force fields for quantum-based molecular dynamics~\cite{behler2007,bartok2010,zhang2018,unke2021,batzner2022,batatia2022,suwa2019}. By replacing repeated electronic calculations with learned force predictions, these models have extended simulations to larger systems and longer times. This strategy has also been extended to condensed-matter lattice Hamiltonians, where learned forces drive the collective dynamics~\cite{chern2026jmmm,chern2026aip,zhang2023,cheng2023b}. Its scalability draws on locality: the force at a site is inferred from its surrounding configuration, allowing the same predictor to be applied throughout a large lattice. Incorporating the relevant microscopic symmetries ensures consistent predictions for physically equivalent local environments. Such models have enabled large-scale studies of electronic phase separation and charge-density-wave (CDW), orbital, and magnetic ordering kinetics~\cite{zhang2021,zhang2022,cheng2023a,ghosh2024,fan2024,tyberg2025}.

An alternative to force-field modeling is to learn a finite-time evolution map directly from order-parameter trajectories, relating a configuration to one at a later time. Such a model advances the field without evaluating microscopic forces and may permit larger effective time steps. In principle, it can also be trained on experimental trajectories without knowledge of the underlying Hamiltonian. Recurrent neural networks (RNNs) provide a natural framework for learning these time sequences, using an evolving hidden state to retain information from previous observations~\cite{elman1990,yu2019,sherstinsky2020}. Architectures ranging from reservoir computing to long short-term memory (LSTM) networks have been used to predict chaotic classical dynamics, molecular trajectories, and quantum evolution~\cite{hochreiter1997,jaeger2004,cho2014,pathak2018,vlachas2018,chattopadhyay2020,tsai2020,banchi2018,flurin2020,lin2021}.

The need for temporal context also has a physical basis. Although microscopic dynamics are Markovian when expressed in terms of a complete state, eliminating electronic and other unobserved degrees of freedom can make the reduced dynamics non-Markovian~\cite{zwanzig1961,mori1965,chorin2000}. Past observations may therefore contain information absent from the present one; under suitable conditions, Takens' theorem shows how delayed observations can reconstruct the underlying dynamical state~\cite{takens1981}. This motivates models that take a short history as input, including recent transformer approaches~\cite{jang2026,herrera2024}.

An RNN instead compresses past observations into a recursively updated hidden state. The joint update of visible variables and hidden states is Markovian, while the visible dynamics alone can retain memory. From a coarse-graining perspective, the hidden states may encode the influence of unresolved microscopic degrees of freedom without identifying them individually~\cite{koppe2019,vlachas2022,durstewitz2023}. Some recurrent models have addressed spatially extended dynamics~\cite{pathak2018,ning2026,dinh2025}, but interacting lattice systems pose a further design question: how can local recurrent memory be coupled through symmetry-aware GNN message passing to predict evolving textures across large lattices?

Here we introduce a graph-recurrent neural network (GRNN) that learns the evolution of spatially distributed observables directly from trajectories, without specifying their microscopic equations of motion. We formulate it as a message-passing extension of a gated recurrent unit (GRU), whose gates regulate how a hidden state incorporates new information and retains its past~\cite{cho2014,chung2014}. On the lattice graph, messages from neighboring observables build a representation of the current spatial texture~\cite{scarselli2009,kipf2017,gilmer2017,sanchezgonzalez2020,fan2026,bakshi2026}. Messages between neighboring hidden states also enter the recurrent update, allowing information accumulated over time to influence nearby sites~\cite{li2016,li2018}. The resulting shared local rule advances the observable and hidden fields together and can be applied to lattices larger than those used for training.

\begin{figure}[t]
 \centering
 \includegraphics[width=0.99\columnwidth]{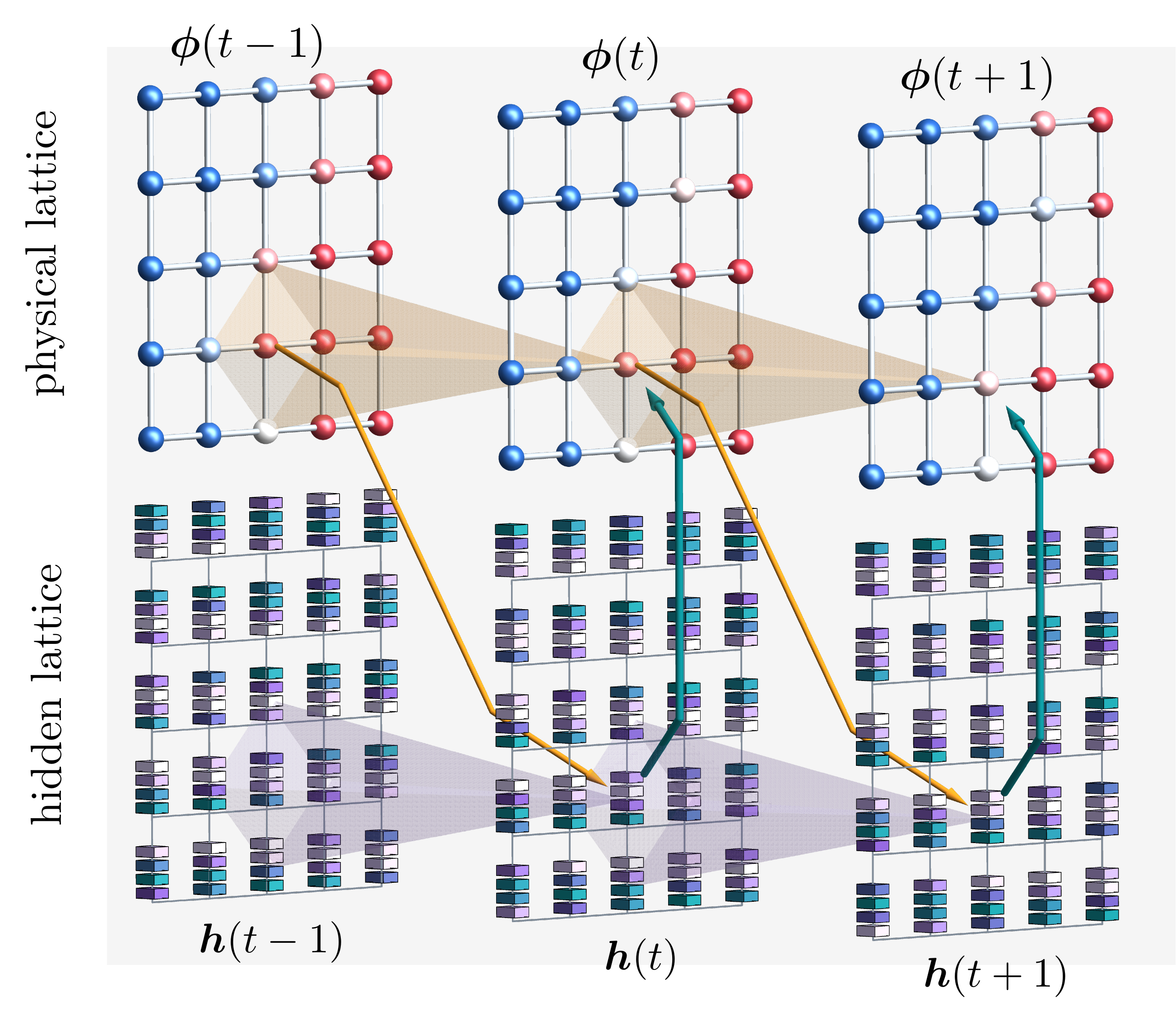}
 \caption{Schematic of the graph-recurrent update. The upper lattices represent successive physical configurations $\bm\phi(t)$, and the lower lattices represent the corresponding site-resolved hidden states $\bm h(t)$. The labels $t-1$, $t$, and $t+1$ denote successive prediction steps. Neighboring sites contribute to both the spatial representation of the physical field and the evolution of the hidden states. The colored arrows indicate the exchange of information between the physical and hidden states. The schematic shows the coupled evolution; the operations within one time step are given in the text.}
 \label{fig:schematic}
\end{figure}

We represent the evolving physical field by $\bm\phi(t)=\{\bm\phi_i(t)\}_{i\in V}$, where $V$ is the set of lattice sites, $i$ labels a site and $\bm\phi_i(t)=(\phi_{i,1}(t),\ldots,\phi_{i,M}(t))$ contains its $M$ order-parameter components. In the GRNN update, $t+1$ denotes the next sampled configuration, separated from $t$ by the prediction interval $\Delta t_{\mathrm{pred}}$ in physical time. At each site we also introduce a hidden vector $\bm h_i(t)$, which evolves alongside the physical field. As illustrated in Fig.~\ref{fig:schematic}, the GRNN learns a coupled update $(\bm\phi(t),\bm h(t))\mapsto(\bm\phi(t+1),\bm h(t+1))$, where $\bm h(t)=\{\bm h_i(t)\}_{i\in V}$. The hidden states are trained to retain information relevant to future evolution that is absent from the instantaneous physical configuration, including the influence of microscopic degrees of freedom excluded from the reduced description. They need not correspond individually to those microscopic variables; their role is to provide an evolving representation of their effects. The combined state can therefore admit a local-in-time update even when the dynamics of $\bm\phi$ alone depend on its history.

The hidden states supply temporal information, but the evolution at a site also depends on the order-parameter texture around it. To incorporate this spatial dependence, we use graph message passing to build an embedding from the current physical configuration. Each successive layer combines information from neighboring sites, allowing the representation at site $i$ to reflect a progressively larger region of the lattice. This gathering of spatial information is illustrated in the upper lattices of Fig.~\ref{fig:schematic}. Starting from $\bm v_i^{(0)}(t)=\bm\phi_i(t)$, we define the mean neighbor features $\overline{\bm v}_i^{(\ell)}(t)=d_i^{-1}\sum_{j\in\mathcal N_i}\bm v_j^{(\ell)}(t)$, where $\mathcal N_i$ is the neighbor set and $d_i=|\mathcal N_i|$. The features are updated according to
\begin{equation}
\bm v_i^{(\ell+1)}(t)=\sigma_{\mathrm G}\!\left[W_{\mathrm{self}}^{(\ell)}\bm v_i^{(\ell)}(t)+W_{\mathrm{nbr}}^{(\ell)}\overline{\bm v}_i^{(\ell)}(t)+\bm b^{(\ell)}\right],
\label{eq:graph_layers}
\end{equation}
where $\sigma_{\mathrm G}$ is the graph-layer activation, chosen here as the Gaussian error linear unit (GELU). The weight matrices and biases are learned and shared across sites. After $N_{\mathrm G}$ layers, $\bm z_i(t)=\bm v_i^{(N_{\mathrm G})}(t)$ encodes the surrounding texture for use in the recurrent update. This embedding is constructed from the configuration at time $t$; information from earlier times enters through the hidden states.

The embedding $\bm z_i(t)$ describes the spatial texture at the current time. The recurrent update combines it with hidden states that carry information from earlier configurations, allowing the model's use of that information to depend on the present texture. Because the history relevant to a site can also originate nearby---for example, as a domain boundary moves toward it---we couple neighboring hidden states along the same lattice bonds. This exchange between hidden states is illustrated in the lower lattices of Fig.~\ref{fig:schematic}. Defining $\overline{\bm h}_i(t)=d_i^{-1}\sum_{j\in\mathcal N_i}\bm h_j(t)$, we construct the reset and update gates from $\bm z_i(t)$, $\bm h_i(t)$, and $\overline{\bm h}_i(t)$:
\begin{equation}
\begin{aligned}
\bm r_i(t)&=\sigma_{\mathrm s}\!\left[W_r\bm z_i(t)+U_r\bm h_i(t)+V_r\overline{\bm h}_i(t)+\bm b_r\right],\\
\bm u_i(t)&=\sigma_{\mathrm s}\!\left[W_u\bm z_i(t)+U_u\bm h_i(t)+V_u\overline{\bm h}_i(t)+\bm b_u\right],
\end{aligned}
\label{eq:recurrent_gates}
\end{equation}
where $\sigma_{\mathrm s}$ is the logistic sigmoid. These gates determine how the stored information contributes to the new hidden state: the reset gate regulates the contribution of $\bm h_i(t)$ to a candidate state, while the update gate controls how much of that candidate is incorporated. Specifically,
\begin{equation}
\begin{aligned}
\widetilde{\bm h}_i(t+1)&=\tanh\!\Bigl[W_h\bm z_i(t)+U_h\bigl(\bm r_i(t)\odot\bm h_i(t)\bigr)\\
&\qquad\qquad+V_h\overline{\bm h}_i(t)+\bm b_h\Bigr],\\
\bm h_i(t+1)&=\bigl[\bm 1-\bm u_i(t)\bigr]\odot\bm h_i(t)+\bm u_i(t)\odot\widetilde{\bm h}_i(t+1),
\end{aligned}
\label{eq:hidden_update}
\end{equation}
where $\odot$ denotes componentwise multiplication. The neighbor contribution enters the candidate directly, so information stored at one site can influence the hidden states nearby and propagate over successive time steps.

The next physical configuration is obtained from the spatial embedding and the updated hidden state:
\begin{equation}
\begin{aligned}
\bm\delta_i(t)&=\mathrm{MLP}\!\left[\bm z_i(t)\oplus\bm h_i(t+1)\right],\\
\bm\phi_i(t+1)&=\mathcal R_\phi\!\left[\bm\phi_i(t),\bm\delta_i(t)\right],
\end{aligned}
\label{eq:physical_readout}
\end{equation}
where $\oplus$ denotes concatenation and MLP denotes a multilayer perceptron. The readout $\mathcal R_\phi$ specifies how a learned change acts on the physical variable and can be adapted to the allowed values and constraints of the order parameter. The complete update uses only $\bm\phi(t)$ and $\bm h(t)$, but the evolution of $\bm\phi$ alone can depend on earlier configurations through the hidden states. In this way, the network combines a local spatial description with a learned, site-dependent representation of memory.

\begin{figure}[t]
 \centering
 \includegraphics[width=0.9\columnwidth]{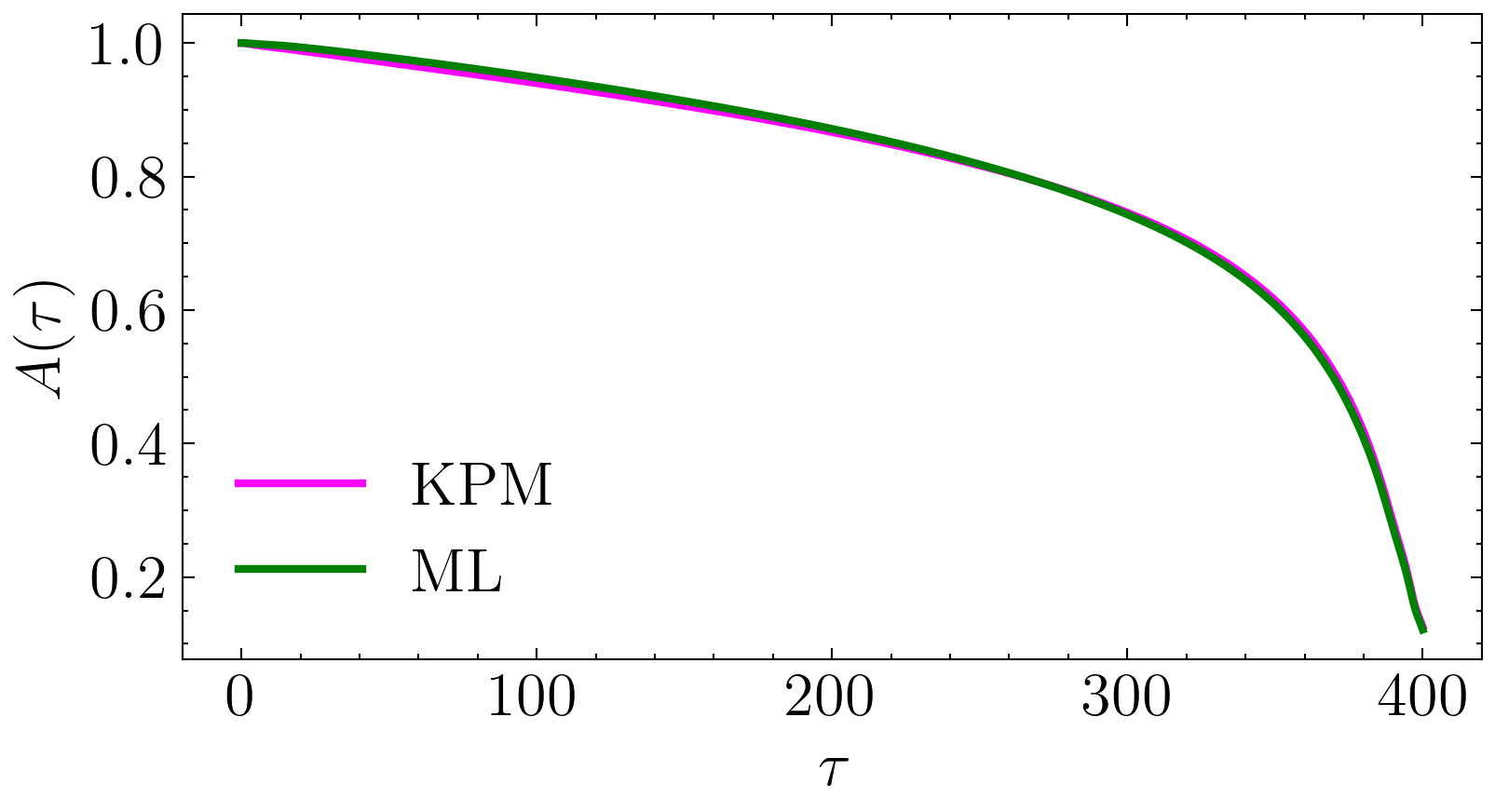}
 \caption{Normalized temporal autocorrelation $A(\tau)$ of the CDW order parameter, comparing GRNN predictions (green) with reference adiabatic dynamics computed using KPM (magenta). The GRNN closely reproduces the decay of temporal correlations over the full interval $0\leq\tau\leq400$.}
 \label{fig:autocorrelation}
\end{figure}

We demonstrate the framework through CDW ordering in the semiclassical Holstein model, which describes itinerant electrons coupled to local lattice distortions~\cite{holstein1959}. For spinless electrons on a square lattice, the Hamiltonian is
\begin{equation}
\begin{aligned}
\mathcal H={}&t_{\rm nn}\sum_{\langle ij\rangle}(c_i^\dagger c_j+\mathrm{H.c.})-g\sum_iQ_i\hat n_i\\
&+\sum_i\left(\frac{P_i^2}{2m}+\frac{k Q_i^2}{2}\right)+ \kappa \sum_{\langle ij\rangle}Q_iQ_j,
\end{aligned}
\label{eq:holstein_hamiltonian}
\end{equation}
where $\hat n_i=c_i^\dagger c_i$, $Q_i$ is a classical lattice displacement, and $P_i$ is its conjugate momentum. The parameters $t_{\rm nn}$ and $g$ specify nearest-neighbor hopping and electron--lattice coupling, while $k$ and $\kappa$ describe the local and intersite elastic interactions.

We adopt the adiabatic approximation, in which the electrons equilibrate to the instantaneous displacement configuration $\bm Q(t)$. The electronic subsystem thus has no independent dynamics, but its density $n_i(t)=\langle\hat n_i\rangle_{\bm Q(t)}$ determines the force acting on the lattice. The classical variables obey
\begin{equation}
m\ddot Q_i+\gamma\dot Q_i=-k  Q_i- \kappa \sum_{j\in\mathcal N_i}Q_j+g n_i(t)+\eta_i(t),
\label{eq:holstein_dynamics}
\end{equation}
where $\gamma$ is the damping coefficient and $\eta_i(t)$ represents thermal noise. Although the electronic Hamiltonian is quadratic at fixed $\bm Q$, the density must be recalculated as the lattice evolves. We generate reference trajectories using a Verlet-type integrator and evaluate the electronic densities and forces at each integration step with the kernel polynomial method (KPM)~\cite{barros2013}.

\begin{figure}[t]
 \centering
 \includegraphics[width=0.99\columnwidth]{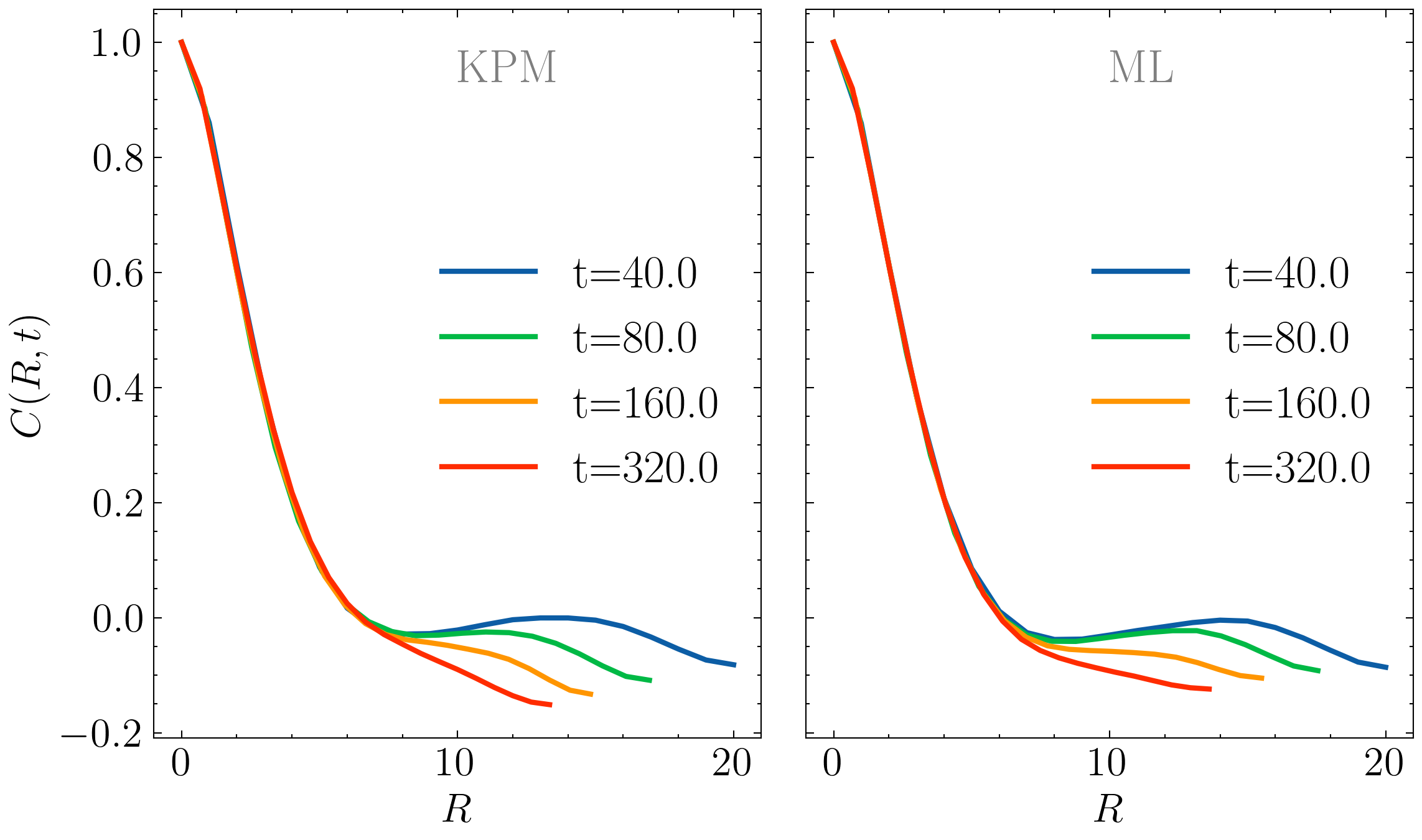}
 \caption{Equal-time CDW correlation function $C(R,t)$ versus scaled separation $R=x/L(t)$ at $t=40$, $80$, $160$, and $320$, obtained from KPM reference dynamics (left) and GRNN predictions (right). The GRNN captures the spatial correlation profiles, while the collapse of curves at different times in both panels demonstrates dynamical scaling.}
 \label{fig:spatial-correlation}
\end{figure}

\begin{figure*}[t]
 \centering
 \includegraphics[width=\textwidth]{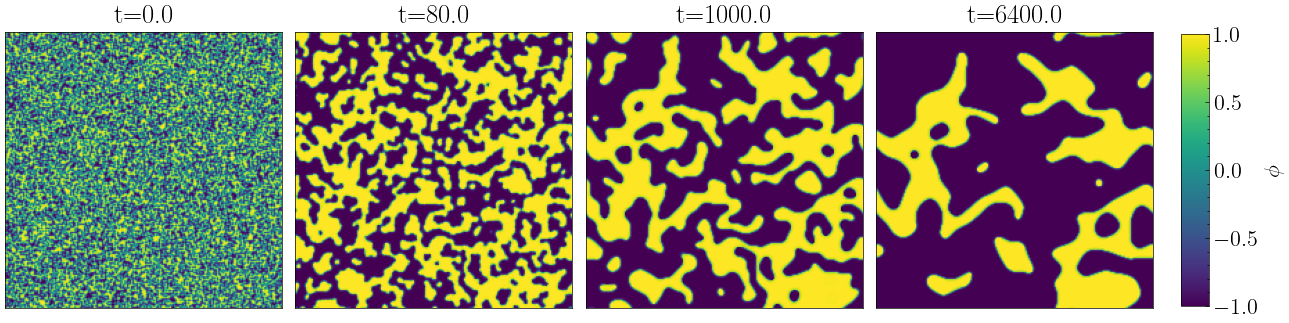}
 \caption{Autoregressive GRNN simulation of CDW domain coarsening on a $200\times200$ lattice at $t=0$, $80$, and $6400$. Color indicates the local order parameter $\phi$ on the scale shown at right, from $-1$ to $1$. The initially disordered field evolves into domains of the two symmetry-related charge-ordered states.}
 \label{fig:large-lattice}
\end{figure*}

At half filling, the square-lattice Holstein model exhibits a strong tendency toward checkerboard charge order~\cite{scalettar1989,noack1991,hohenadler2019}. Its microscopic origin is particularly transparent for nearest-neighbor hopping, where perfect nesting of the half-filled Fermi surface enhances the charge response to a checkerboard modulation. The resulting divergent charge response makes the uniform state unstable to a staggered lattice distortion for arbitrarily weak nonzero $g$ at zero temperature in the classical lattice limit~\cite{weber2018}. This distortion opens an electronic gap and produces alternating high- and low-density sites. Interchanging the two sublattices gives an equivalent ordered state, so the checkerboard CDW breaks a discrete $\mathbb Z_2$ symmetry and undergoes an Ising transition~\cite{noack1991,weber2018}.

During phase ordering, regions of these two checkerboard patterns coexist and evolve through domain-wall motion. To describe this texture, we define the neighbor-averaged density $\overline n_i(t)=\frac14\sum_{j\in\mathcal N_i}n_j(t)$ and introduce the local order parameter
\begin{equation}
\phi_i(t)=(-1)^{x_i+y_i}\left[n_i(t)-\overline n_i(t)\right].
\label{eq:local_cdw}
\end{equation}
Here $(x_i,y_i)$ are the integer lattice coordinates of site $i$. The density contrast measures the local charge modulation, while the sublattice factor removes its checkerboard alternation. The resulting field has approximately uniform sign within each domain, with opposite signs distinguishing the two ordered configurations.

The evolution of this emergent field is the quantity learned by the GRNN. In the adiabatic dynamics, the electronic density is an instantaneous function of the lattice displacements, $n_i=n_i(\{Q\})$. Since $\dot Q_k=P_k/m$, the chain rule gives
\begin{equation}
\dot{\phi}_i(t)=(-1)^{x_i+y_i}\sum_k\left[\frac{\partial n_i}{\partial Q_k}-\frac14\sum_{j\in\mathcal N_i}\frac{\partial n_j}{\partial Q_k}\right]\frac{P_k(t)}{m},
\label{eq:cdw_formal_dynamics}
\end{equation}
where the derivatives are evaluated at the instantaneous lattice configuration. This equation is not closed in $\bm\phi$: its right-hand side depends on the microscopic displacements and momenta, which are not determined by the instantaneous order-parameter field. Thus, even beyond the adiabatic approximation, retaining only $\bm\phi$ produces effective non-Markovian dynamics; for example, states with the same displacements but different momenta can have the same $\bm\phi$ and evolve differently. The GRNN addresses this missing information through recurrent states that encode the influence of unresolved lattice motion, coupled through graph messages to account for spatial interactions. A central test of this learned description is whether it reproduces the temporal correlations of the emergent field. For the half-filled model, the numerical parameters and reference-trajectory protocol are summarized in Appendix~\ref{app:holstein}.


The dataset comprises $30$ independent trajectories, divided into $16$ for training, $4$ for validation, and $10$ for testing, with complete trajectories kept separate across these subsets. We use three graph layers with feature dimensions $1\to128\to96\to64$ and a readout MLP with four hidden layers and GELU activations. The bounded readout, specified in Appendix~\ref{app:training}, uses $\epsilon=10^{-5}$ and a trainable $\alpha$ initialized to $0.1$. Hidden states are initialized to zero and conditioned on reference configurations during an initial burn-in period. Training minimizes the trajectory mean squared error using truncated backpropagation through time and AdamW, with cosine annealing and gradient clipping. After burn-in, reference inputs are used with probability $0.2$ during training; evaluation is fully autoregressive. The correlation benchmarks below use the held-out test trajectories.

We first examine the normalized single-site autocorrelation function, $A(\tau)=\langle\phi_i(t_0)\phi_i(t_0+\tau)\rangle/\langle\phi_i(t_0)^2\rangle$, where $\langle\cdots\rangle$ denotes an average over lattice sites $i$ and reference times $t_0$ for which both configurations are available. Here and below, $t$, $t_0$, and $\tau$ denote physical times. This quantity probes the temporal persistence of local CDW order and provides a direct test of the time sequences learned by the recurrent model. As shown in Fig.~\ref{fig:autocorrelation}, the GRNN closely reproduces the KPM reference results over the displayed time-lag range. While this agreement establishes the accuracy of local temporal correlations, characterizing collective phase ordering also requires examining how correlations develop across the lattice.

We therefore turn to the normalized equal-time spatial correlation function $C(R,t)$, which probes the organization of the CDW field into growing domains. Figure~\ref{fig:spatial-correlation} compares the KPM reference dynamics and GRNN predictions at several stages of the evolution, with the separation $x$ rescaled by the instantaneous correlation length $L(t)$, so that the horizontal axis is $R=x/L(t)$. The results demonstrate two complementary features. First, the GRNN reproduces the spatial correlation profiles of the microscopic dynamics, including their decay and weak anticorrelations at larger scaled separations. Second, the curves at different times collapse onto a common profile in both panels, consistent with dynamical scaling: the evolving domain structure is statistically self-similar when lengths are measured in units of $L(t)$. Beyond capturing local temporal behavior, the GRNN thus reproduces the collective spatial correlations and scaling structure characteristic of CDW coarsening.

Finally, we apply the trained GRNN to a $200\times200$ lattice, demonstrating its ability to simulate extended CDW textures, as shown in Fig.~\ref{fig:large-lattice}. The evolving domain patterns are qualitatively consistent with those obtained in previous ML force-field simulations using descriptor-based and graph neural network models~\cite{cheng2023a,fan2026}. These established features of CDW phase ordering provide a physical context for the large-scale demonstration, whose primary purpose here is to illustrate the scalability of the learned graph-recurrent dynamics. Combined with the quantitative temporal and spatial correlation benchmarks above, these results support the GRNN as an effective reduced description of collective ordering.

The dynamical-scaling analysis gives an algebraic growth law for the characteristic domain size, $L(t)\sim t^\alpha$, with $\alpha<1/2$, slower than the Allen--Cahn law for a nonconserved scalar order parameter~\cite{allen1979}. This sub-Allen--Cahn growth is consistent with previous ML force-field simulations of CDW coarsening~\cite{cheng2023a}. Its physical origin is a kinetic constraint imposed by microscopic electron-number conservation. Although staggered CDW order is nonconserved, nearly binary site occupations tie local domain-wall displacement to charge rearrangement; consequently, a wall segment generally cannot advance independently but must move cooperatively with compatible segments. This suppresses the independent curvature-driven motion underlying Allen--Cahn growth, while thermally induced partial occupations at the walls can relax the constraint and allow a crossover toward the Allen--Cahn regime~\cite{jang2026kinks,bora2026cooperative}.

In summary, we have introduced a GRNN framework that combines local recurrent memory with graph message passing to learn the evolution of spatially distributed observables. Its hidden states encode information absent from the instantaneous physical field, while spatial coupling allows this information to influence the dynamics of neighboring sites. Learning finite-time evolution maps already allows the system to advance over intervals larger than the microscopic integration step; extending the prediction interval further could provide additional gains in computational efficiency. Because training requires trajectories of the selected observables rather than explicit microscopic equations, the framework is also suited to data-driven modeling from experimental measurements, where the underlying Hamiltonian or the relevant microscopic degrees of freedom may be incompletely known.

A particularly compelling direction is the extension to nonadiabatic dynamics. Although the present adiabatic setting already involves an emergent field with unresolved lattice motion, the electrons remain equilibrated to the instantaneous lattice configuration. In nonadiabatic or time-dependent Hartree--Fock dynamics, the electronic sector has its own evolution, and its feedback on the order parameter can depend on electronic excitations and coherences that are not determined by the instantaneous order. Learning only the order-parameter evolution would therefore require the recurrent states to encode the influence of this much larger unresolved sector. Such applications may demand larger hidden-state dimensions or more expressive memory architectures, although the required capacity is governed by the information relevant to collective evolution rather than the total number of microscopic variables. Extending the GRNN in this direction would provide a route to reduced dynamical models that retain electronic memory while avoiding explicit propagation of the full electronic state.

\begin{acknowledgments}
This work was supported by the U.S. Department of Energy, Office of Science, Office of Basic Energy Sciences, under Award No. DE-SC0020330.   The authors acknowledge Research Computing at the University of Virginia for providing computational resources and technical support.
\end{acknowledgments}

\onecolumngrid
\raggedbottom 
\section*{Appendix}
\appendix

\section{Graph-recurrent update}
\label{app:architecture}

At each lattice site $i\in V$, the GRNN associates the physical observable $\bm\phi_i(t)=(\phi_{i,1}(t),\ldots,\phi_{i,M}(t))$ with a multicomponent hidden state $\bm h_i(t)$. The observable and hidden fields, $\bm\phi(t)=\{\bm\phi_i(t)\}_{i\in V}$ and $\bm h(t)=\{\bm h_i(t)\}_{i\in V}$, evolve together through the coupled update $(\bm\phi(t),\bm h(t))\mapsto(\bm\phi(t+1),\bm h(t+1))$. The hidden states retain information from earlier configurations, while graph messages couple neighboring sites. All sites share the same learned update rule.

\begin{figure}
 \centering
 \resizebox{\linewidth}{!}{%
 \begin{tikzpicture}[
  >=Latex,
  node distance=6mm,
  every node/.style={font=\sffamily\small,align=center},
  state/.style={draw=blue!70!black,fill=blue!6,rounded corners=2pt,minimum height=13mm,minimum width=21mm},
  spatial/.style={draw=orange!85!black,fill=orange!7,rounded corners=2pt,minimum height=13mm,minimum width=31mm},
  memory/.style={draw=violet!80!black,fill=violet!7,rounded corners=2pt,minimum height=13mm,minimum width=29mm},
  head/.style={draw=teal!75!black,fill=teal!7,rounded corners=2pt,minimum height=13mm,minimum width=27mm},
  bound/.style={draw=black!70,fill=black!4,rounded corners=2pt,minimum height=13mm,minimum width=30mm},
  flow/.style={-{Latex[length=2.2mm]},line width=0.75pt},
  upper/.style={flow,draw=orange!85!black},
  lower/.style={flow,draw=violet!80!black}]
  \node[state] (phi) {$\bm\phi_i(t)$};
  \node[spatial,right=of phi] (gnn) {graph layers\\$\bm v_i^{(\ell+1)}=\sigma_{\mathrm G}\!\left(W_{\rm self}^{(\ell)}\bm v_i^{(\ell)}+W_{\rm nbr}^{(\ell)}\overline{\bm v}_i^{(\ell)}+\bm b^{(\ell)}\right)$};
  \node[spatial,right=of gnn,minimum width=17mm] (z) {$\bm z_i(t)$};
  \node[head,right=of z] (mlp) {readout MLP\\$\bm\delta_i(t)=\mathrm{MLP}(\bm z_i\oplus\bm h_i^{+})$};
  \node[bound,right=of mlp] (res) {physical update\\$\mathcal R_\phi(\bm\phi_i,\bm\delta_i)$};
  \node[state,right=of res] (nextphi) {$\bm\phi_i(t+1)$};
  \node[memory,below=17mm of gnn] (oldh) {local and neighbor memory\\$\bm h_i(t),\ \overline{\bm h}_i(t)$};
  \node[memory,right=12mm of oldh] (gru) {graph-coupled GRU\\$\bm h_i^{+}=\mathrm{GRU}_{\rm graph}(\bm z_i,\bm h_i,\overline{\bm h}_i)$};
  \node[memory,right=12mm of gru,minimum width=20mm] (newh) {$\bm h_i^{+}$\\$=\bm h_i(t+1)$};
  \draw[upper] (phi) -- (gnn);
  \draw[upper] (gnn) -- (z);
  \draw[upper] (z) -- (mlp);
  \draw[flow,draw=teal!75!black] (mlp) -- (res);
  \draw[flow] (res) -- (nextphi);
  \draw[lower] (oldh) -- (gru);
  \draw[lower] (gru) -- (newh);
  \draw[upper] (z.south) |- ($(gru.north)+(0,3mm)$) -- (gru.north);
  \draw[flow,draw=teal!75!black] (newh.north) |- ($(mlp.south)+(0,-4mm)$) -- (mlp.south);
  \draw[flow,draw=black!75] (phi.north) -- ++(0,11mm) -| (res.north);
  \node[font=\sffamily\scriptsize,above=10mm of z] {current field bypass};
  \draw[lower] (newh.south) -- ++(0,-7mm) -- ++(28mm,0) node[right,font=\sffamily\scriptsize] {memory passed to next step};
 \end{tikzpicture}}
 \caption{One-step GRNN computation. Graph layers aggregate neighboring physical features into the embedding $\bm z_i(t)$; the graph-coupled GRU combines this embedding with local and neighboring hidden states to produce $\bm h_i^{+}\equiv\bm h_i(t+1)$. The readout MLP uses $\bm z_i(t)$ and the updated memory to predict $\bm\delta_i(t)$, which enters the physical update map $\mathcal R_\phi$. The upper bypass carries the present physical field directly to this map. Spatial and recurrent aggregation are indicated by orange and violet paths, respectively. All learned parameters are shared across lattice sites.}
 \label{fig:appendix-flow}
\end{figure}
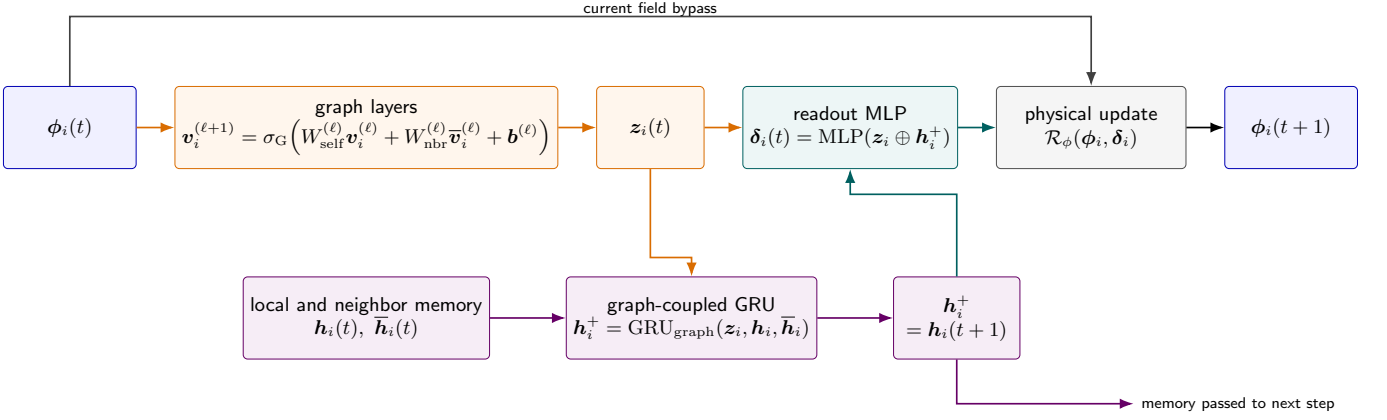

This appendix gives a more detailed view of this update in Eqs.~\eqref{eq:graph_layers}--\eqref{eq:physical_readout}. Figure~\ref{fig:appendix-flow} follows one prediction step. Its upper path constructs spatial features from the present order-parameter configuration and advances the physical field; its lower path updates the site-resolved recurrent memory. The diagram uses the notation of the main text throughout: $t+1$ is the next \emph{sampled} time, a physical interval $\Delta t_{\mathrm{pred}}$ after $t$. Thus $\bm h_i(t+1)$ enters the prediction of $\bm\phi_i(t+1)$ and is also carried into the following step.

For clarity, the neighbor averages are $\overline{\bm v}_i^{(\ell)}=d_i^{-1}\sum_{j\in\mathcal N_i}\bm v_j^{(\ell)}$ and $\overline{\bm h}_i=d_i^{-1}\sum_{j\in\mathcal N_i}\bm h_j$, with $d_i=|\mathcal N_i|$. Defining $\bm a_i(t)=\bm z_i(t)\oplus\bm h_i(t)\oplus\overline{\bm h}_i(t)$ allows the two gates in Eq.~\eqref{eq:recurrent_gates} to be written compactly as $\bm r_i=\sigma_{\rm s}(A_r\bm a_i+\bm b_r)$ and $\bm u_i=\sigma_{\rm s}(A_u\bm a_i+\bm b_u)$. The matrices $A_r$ and $A_u$ simply concatenate the $W$, $U$, and $V$ matrices of the main text; this notation does not change the architecture. The candidate and updated memories are
\begin{equation}
\begin{aligned}
\widetilde{\bm h}_i^{+}&=\tanh\!\left[W_h\bm z_i+U_h(\bm r_i\odot\bm h_i)+V_h\overline{\bm h}_i+\bm b_h\right],\\
\bm h_i^{+}&=(\bm1-\bm u_i)\odot\bm h_i+\bm u_i\odot\widetilde{\bm h}_i^{+},
\end{aligned}
\label{eq:app-memory}
\end{equation}
where quantities without an explicit time argument are evaluated at $t$ and $\bm h_i^{+}=\bm h_i(t+1)$. The reset gate acts on the \emph{local} previous memory in the candidate, while the neighboring memory contributes through the separate $V_h$ term. An update at one site can therefore be influenced by memories stored at adjacent sites even if the current local observable happens to be the same.

With nearest-neighbor graph layers, $\bm z_i(t)$ depends on observables out to $N_{\mathrm G}$ graph hops at the same time. Repeated recurrent steps allow information in the hidden field to move farther across the lattice. The readout then gives $\bm\delta_i=\mathrm{MLP}(\bm z_i\oplus\bm h_i^{+})$ and $\bm\phi_i(t+1)=\mathcal R_\phi[\bm\phi_i(t),\bm\delta_i]$. The map $\mathcal R_\phi$ can be chosen to respect the range or geometric constraints of the physical field. These operations make the augmented $(\bm\phi,\bm h)$ update local in time without requiring the observable field alone to be Markovian.

\section{Holstein reference dynamics and CDW observables}
\label{app:holstein}

We use the spinless square-lattice Hamiltonian of Eq.~\eqref{eq:holstein_hamiltonian}. At any fixed displacement configuration $\bm Q$, the electronic part is a quadratic hopping problem with a site-dependent potential $-gQ_i$; its density $n_i[\bm Q]=\langle\hat n_i\rangle_{\bm Q}$ enters the force on the classical coordinate. Writing $P_i=m\dot Q_i$, the adiabatic stochastic dynamics are equivalently
\begin{equation}
\dot Q_i=\frac{P_i}{m},\qquad \dot P_i=-kQ_i-\kappa\sum_{j\in\mathcal N_i}Q_j+g\,n_i[\bm Q]-\frac{\gamma}{m}P_i+\eta_i(t).
\label{eq:app-holstein-force}
\end{equation}
The force must be reevaluated as $\bm Q$ changes. We compute its electronic contribution using the kernel polynomial method (KPM), which approximates spectral functions of the instantaneous one-electron Hamiltonian with a truncated Chebyshev expansion and evaluates local densities without full diagonalization~\cite{weisse2006,barros2013}. For the reference trajectories, we retain $1000$ Chebyshev moments and use $100$ colors in standard calculations or $400$ in higher-precision evaluations (Table~\ref{tab:app-reference}). The classical evolution is integrated with a Verlet-type scheme.

\begin{table}
\caption{Model and reference-trajectory settings. The damping value refers to the integrator setting; no numerical value for $\gamma$ in Eq.~\eqref{eq:app-holstein-force} is inferred from it.}
\label{tab:app-reference}
\begin{ruledtabular}
\begin{tabular}{ll}
Quantity & Value \\
Hopping $t_{\rm nn}$; coupling $g$ & $-1$; $3.5$ \\
Elastic constants $k$, $\kappa$; mass $m$ & $1$, $0.18$; $5$ \\
Filling; temperature $k_{\rm B}T$ & $\overline n=1/2$; $0.3$ \\
Reference lattice & $40\times40$ sites \\
Integration step $\delta t$; integrator damping setting & $0.04$; $0.2$ \\
Steps per trajectory; final time & $10^4$; $400$ \\
KPM Chebyshev order & $1000$ \\
KPM colors, standard / higher precision & $100$ / $400$ \\
Snapshot spacing; prediction interval $\Delta t_{\rm pred}$ & $20$ steps; $0.8$ \\
Independent trajectories, train / validation / test & $16$ / $4$ / $10$ \\
\end{tabular}
\end{ruledtabular}
\end{table}

\begin{figure}[t]
 \centering
 \includegraphics[width=0.65\linewidth]{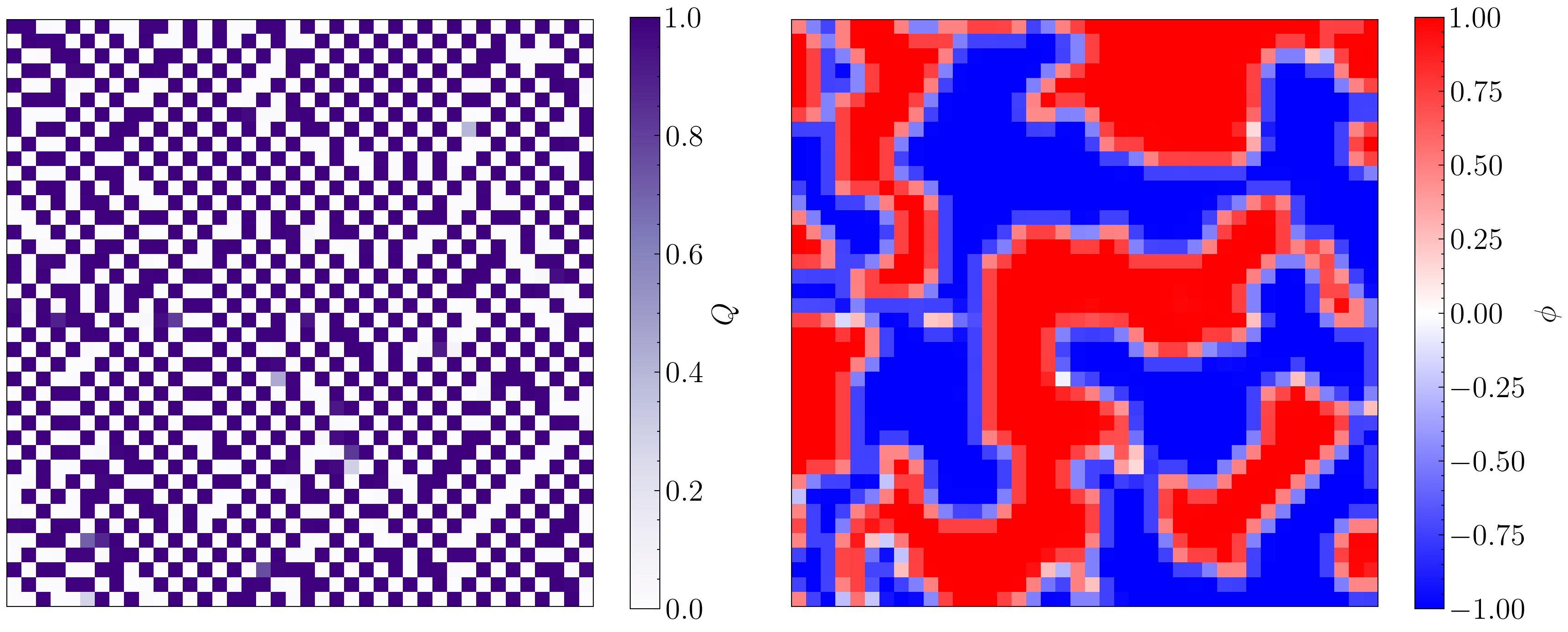}
 \caption{Microscopic lattice pattern and emergent Ising-like CDW field in a representative $40\times40$ configuration. Left: displacement $Q_i$, whose alternating pattern tracks the local charge modulation through electron--lattice coupling. Right: staggered neighbor-density contrast $\phi_i=(-1)^{x_i+y_i}(n_i-\frac14\sum_{j\in\mathcal N_i}n_j)$ evaluated for the same configuration. The staggered factor removes the checkerboard oscillation, so red and blue regions mark the two symmetry-related CDW domains; intermediate values occur near domain walls. The color bars show the values plotted in each panel.}
 \label{fig:app-cdw-map}
\end{figure}

For nearest-neighbor hopping on the square lattice, the electronic dispersion and nesting relation are
\begin{equation}
\varepsilon_{\bm k}=2t_{\rm nn}(\cos k_x+\cos k_y),\qquad \varepsilon_{\bm k+\bm q_{\rm CDW}}=-\varepsilon_{\bm k},
\label{eq:app-dispersion-nesting}
\end{equation}
where $\bm q_{\rm CDW}=(\pi,\pi)$ and the lattice spacing is set to unity. The half-filled Fermi surface is therefore perfectly nested at the checkerboard ordering wave vector. This nesting underlies the instability toward staggered charge and lattice order discussed in the main text~\cite{weber2018}. The two symmetry-related checkerboard patterns differ by an exchange of the sublattices. A local density contrast $n_i-\frac14\sum_{j\in\mathcal N_i}n_j$ emphasizes the modulation while suppressing a uniform density background. Multiplying by $(-1)^{x_i+y_i}$ then removes the rapid checkerboard sign, giving the smooth Ising-like field $\phi_i$ of Eq.~\eqref{eq:local_cdw}. For a fully polarized ideal checkerboard with occupations $1$ and $0$, this definition yields $\phi_i=+1$ throughout one domain and $\phi_i=-1$ throughout the translated domain. Across an interface it interpolates between the two signs. Figure~\ref{fig:app-cdw-map} illustrates how the strongly alternating microscopic pattern becomes a field with large, readily identifiable domains.

\begin{figure}
 \centering
 \includegraphics[width=0.7\linewidth]{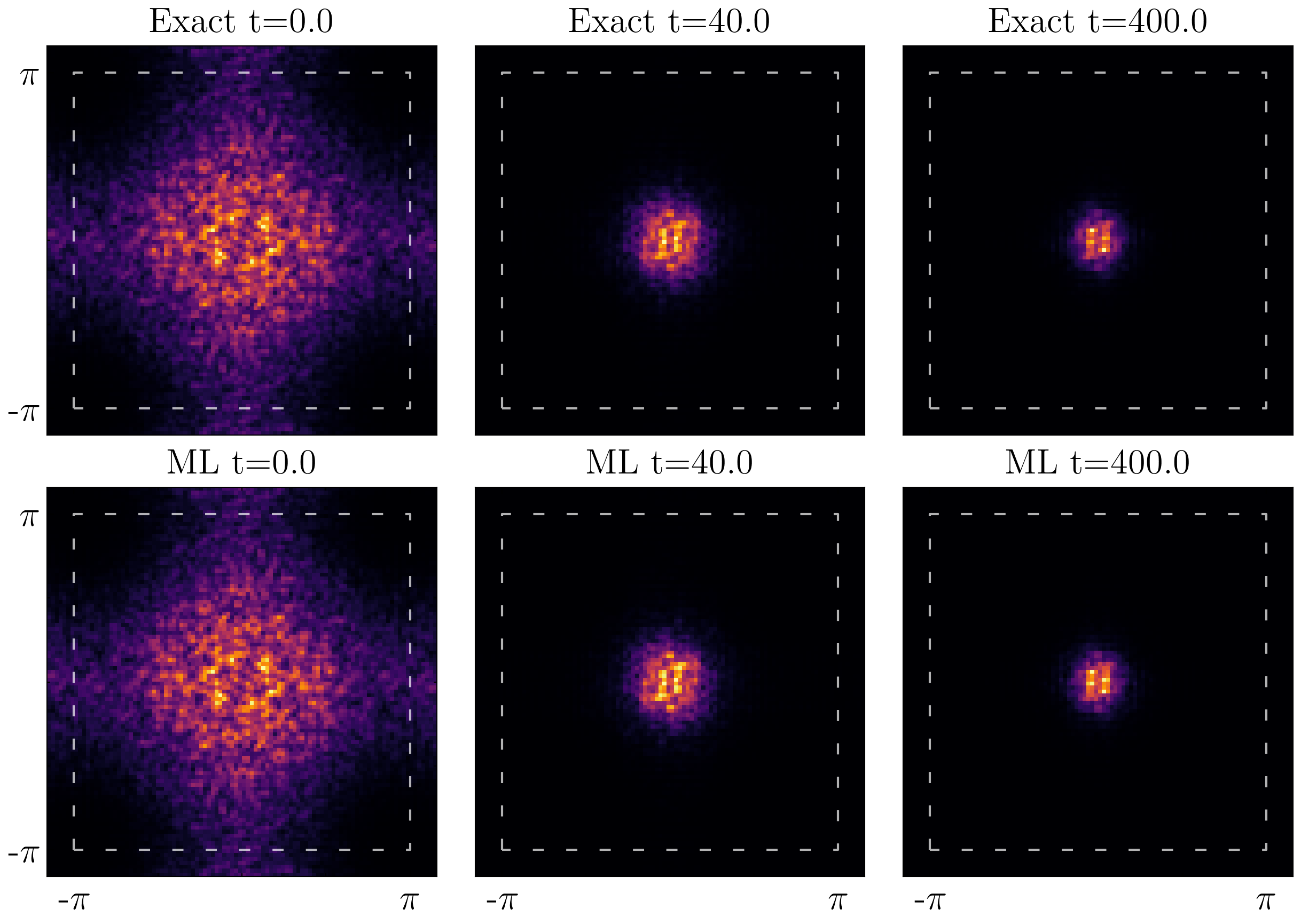}
 \caption{Time-dependent structure factor $S(\bm q,t)$ of the staggered CDW field on a $40\times40$ lattice. The upper row shows KPM reference trajectories and the lower row shows autoregressive GRNN predictions at $t=0$, $40$, and $400$ (left to right). Momentum is measured relative to the checkerboard ordering wave vector, so the ordering peak of the staggered field is centered at $\bm q=\bm0$. Its narrowing as time increases reflects growing CDW domains; the GRNN reproduces the corresponding momentum-space pattern. Dashed squares outline the first Brillouin zone.}
 \label{fig:app-structure-factor}
\end{figure}

The equal-time structure factor provides a complementary momentum-space view of coarsening. For the staggered field we define
\begin{equation}
\phi_{\bm q}(t)=\sum_i\phi_i(t)e^{-i\bm q\cdot\bm r_i},\qquad S(\bm q,t)=\frac{1}{N}\left\langle\left|\phi_{\bm q}(t)\right|^2\right\rangle,
\label{eq:app-structure-factor}
\end{equation}
where $N$ is the number of lattice sites and the brackets indicate an average over independent trajectories. Because the checkerboard factor is already included in $\phi_i$, the CDW peak occurs at $\bm q=\bm0$ in this representation; in the density structure factor the corresponding peak is at $(\pi,\pi)$. Figure~\ref{fig:app-structure-factor} compares the KPM and GRNN maps at three times. The broad initial peak becomes narrower as spatial correlations extend to larger distances, and the GRNN reproduces the evolution of its momentum-space shape.

\section{Kernel Polynomial Method for Reference Data Generation}
\label{app:kpm}

The reference trajectories used for training and evaluating the
GRNN are generated by integrating the Langevin equations of
motion for the classical lattice degrees of freedom, with the
electronic forces evaluated using the kernel polynomial method
(KPM)~\cite{weisse2006,barros2013}. This approach avoids the computationally expensive
diagonalization of the electronic Hamiltonian by approximating
functions of the single-particle Hamiltonian through a
Chebyshev polynomial expansion.

For a given configuration of lattice displacements, the
single-particle Hamiltonian $\mathcal{H}$ is rescaled such
that its spectrum lies within the interval $[-1,1]$:
\begin{equation}
    \widetilde{\mathcal{H}}
    = \frac{\mathcal{H}-b}{a},
\end{equation}
where $a$ and $b$ are determined by the spectral bounds.
A function of the Hamiltonian can then be approximated as
\begin{equation}
    f(\mathcal{H})
    \approx \sum_{m=0}^{M-1}
    c_m T_m(\widetilde{\mathcal{H}}),
\end{equation}
where $T_m$ denotes the Chebyshev polynomial of order $m$,
$M$ is the expansion order, and $c_m$ are the corresponding
expansion coefficients, including any damping factors.
The Chebyshev polynomials are evaluated recursively through
\begin{equation}
    T_{m+1}(\widetilde{\mathcal{H}})
    = 2\widetilde{\mathcal{H}}T_m(\widetilde{\mathcal{H}})
    - T_{m-1}(\widetilde{\mathcal{H}}).
\end{equation}

For noninteracting electrons coupled to classical lattice
displacements, the electronic force acting on the displacement
$x_i$ can be expressed in terms of the electronic free energy
$F_{\mathrm{el}}$ as
\begin{equation}
    F_i^{\mathrm{el}}
    = -\frac{\partial F_{\mathrm{el}}}{\partial x_i}
    = -\mathrm{Tr}\left[
        f_{\mathrm{FD}}(\mathcal{H})
        \frac{\partial\mathcal{H}}{\partial x_i}
    \right],
\end{equation}
where $f_{\mathrm{FD}}$ is the Fermi--Dirac distribution and
the trace is over the single-particle Hilbert space.
These forces are evaluated using KPM and incorporated into
the Langevin equations governing the lattice dynamics.

Because the Chebyshev recurrence requires only sparse
matrix-vector multiplications, the computational cost scales
linearly with the number of lattice sites for fixed expansion
order and stochastic sampling parameters\cite{weisse2006}. The resulting
microscopic trajectories are subsequently mapped onto the
local charge-density-wave order parameter, providing the
training data and reference benchmarks for the GRNN.

\section{Network implementation and training for the Holstein demonstration}
\label{app:training}

For the Holstein demonstration, the input is one scalar $\phi_i$ per lattice site. Three graph aggregation layers map channel dimensions $1\to128\to96\to64$, with GELU as the graph-layer nonlinearity. Their output $\bm z_i$ enters the graph-coupled GRU described above. A sitewise MLP with four hidden layers and GELU activations reads out a scalar increment $\delta_i$ from $\bm z_i\oplus\bm h_i^{+}$. The same graph, recurrent, and readout parameters are used at every site, so their count does not increase with lattice area.

The CDW field defined in Eq.~\eqref{eq:local_cdw} is bounded by $-1\leq\phi_i\leq1$. For this scalar observable ($M=1$), we specialize the general physical readout to
\begin{equation}
\mathcal R_\phi(\phi,\delta)=\tanh\!\left[\operatorname{arctanh}\!\bigl(\operatorname{clip}(\phi,-1+\epsilon,1-\epsilon)\bigr)+\alpha\delta\right],
\label{eq:bounded_readout}
\end{equation}
where $\epsilon=10^{-5}$ avoids the singular endpoints and $\alpha$ is a learned scale initialized to $0.1$. This form applies the learned change in the transformed coordinate $\operatorname{arctanh}(\phi)$ while keeping the resulting field within its allowed range. It is a choice specific to the bounded CDW observable; the general GRNN update does not require this readout form.

\begin{table}[H]
\caption{GRNN architecture and training settings used for the Holstein demonstration.}
\label{tab:app-network}
\begin{ruledtabular}
\begin{tabular}{ll}
Component or procedure & Setting \\
Observable channels & $M=1$ \\
Graph layers and channel sequence & $3$; $1\to128\to96\to64$ \\
Graph activation and aggregation & GELU; mean over nearest neighbors \\
Recurrent unit & GRU with local and neighbor hidden inputs \\
Readout & MLP with four hidden layers, GELU \\
Bounded residual parameters & $\epsilon=10^{-5}$; $\alpha$ initialized at $0.1$ \\
Teacher forcing after burn-in & probability $p=0.2$ \\
Optimization & AdamW, cosine learning-rate annealing \\
Sequence gradients & TBPTT, gradient clipping \\
Held-out evaluation & Autoregressive after burn-in \\
\end{tabular}
\end{ruledtabular}
\end{table}

Each trajectory is initialized with $\bm h_i=\bm0$ at every site. During an initial burn-in segment, the reference configurations are supplied as inputs so that the hidden states can encode information from the trajectory before the model is trained to predict freely; these burn-in steps are excluded from the loss. After burn-in, training uses a mixture of reference and predicted configurations: at each step, the reference field is supplied with probability $p=0.2$, and otherwise the model's previous prediction is used. This exposes the model to its own prediction errors during training while retaining some reference inputs. During validation and testing, reference configurations are used only during burn-in; subsequent evolution is fully autoregressive. Thus the correlations reported in the main text and Fig.~\ref{fig:app-structure-factor} assess extended predictions on held-out trajectories, rather than repeated one-step predictions conditioned on exact intermediate fields.

Let $\phi^{(a)}_i(t)$ denote the reference field for trajectory $a$ and $\widehat\phi^{(a)}_i(t+1)$ the corresponding GRNN prediction. The trajectory mean-squared-error objective, evaluated after burn-in, is
\begin{equation}
\mathcal L_{\rm traj}=\frac{1}{BN}\sum_{a=1}^{B}\frac{1}{|\mathcal T_a|}\sum_{t\in\mathcal T_a}\sum_{i=1}^{N}\left[\widehat\phi_i^{(a)}(t+1)-\phi_i^{(a)}(t+1)\right]^2,
\label{eq:app-loss}
\end{equation}
where $B$ is the number of trajectories in the batch, $N$ is the number of lattice sites, and $\mathcal T_a$ contains the prediction times retained after burn-in. The time average is taken separately for each trajectory before averaging over the batch, so each trajectory contributes equally to the objective.

Training uses truncated backpropagation through time (TBPTT) to limit the memory and computation required for long sequences. The forward evolution proceeds continuously across successive temporal chunks: the predicted field and hidden states at the end of one chunk provide the initial values for the next. At each chunk boundary, both are detached from the computational graph. Their numerical values therefore carry the trajectory history forward, while gradients propagate only within each chunk. Gradients are accumulated across chunks over the training interval before updating the parameters. We use AdamW with cosine learning-rate annealing and gradient clipping. Complete trajectories, rather than individual snapshots, are assigned to the training, validation, and test sets, preventing configurations from the same time sequence from appearing in different subsets.

\section{Ablation Study of Candidate Models} \label{app:ablation}

\begin{figure}[t]
    \centering
    \includegraphics[width=0.8\linewidth]{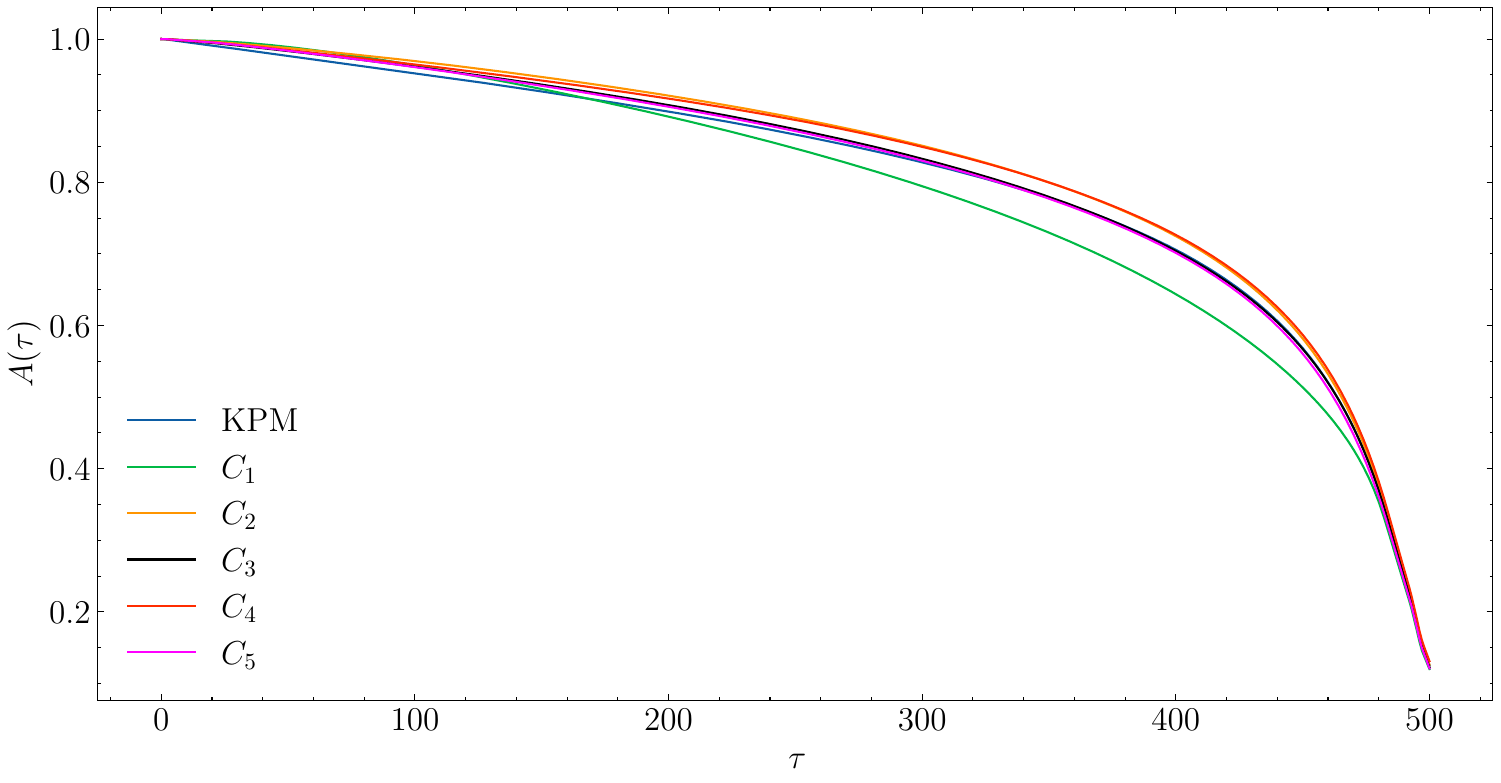}
    \caption{
        Ablation study of the graph-recurrent neural network
        architecture. Temporal autocorrelation functions
        $A(\tau)$ predicted by five candidate models
        ($C_1$--$C_5$) are compared with the KPM reference.
        The candidates differ in their use of recurrent memory
        and spatially aggregated hidden states, as summarized
        in Table~\ref{tab:RMSE}. The full GRNN architecture
        ($C_5$) and the model with spatially coupled recurrent
        gates ($C_3$) exhibit the closest agreement with
        the reference dynamics.
    }
    \label{fig:Abelation}
\end{figure}

To assess the importance of the architectural components of the proposed graph-recurrent neural network (GRNN), we perform an ablation study involving five candidate models. In particular, we investigate the role of recurrent memory and the incorporation of spatially aggregated hidden states into the recurrent updates. The models are evaluated by comparing the temporal autocorrelation functions obtained from autoregressive simulations against the reference results computed using the kernel polynomial method (KPM). 

To ensure a consistent comparison, all models employ the same GNN backbone, consisting of three message-passing layers with feature dimensions $1\rightarrow128\rightarrow96\rightarrow64$. For models containing recurrent units, the hidden-state dimension is fixed at 96. The training procedure is also kept identical across all configurations, using sequences of 50 time steps, a burn-in period of 20 steps, and a teacher-forcing probability of 0.2. All models are trained for 200 epochs with a batch size of 4 and an initial learning rate of $2\times10^{-4}$. For the memoryless model, the burn-in procedure does not initialize any recurrent hidden state. 

The first candidate, denoted by $C_1$, consists of a conventional GNN without recurrent connections. This model performs spatial message passing and predicts the evolution of the order parameter autoregressively. The second candidate, $C_2$, combines the GNN backbone with a conventional GRU, whose gates and candidate hidden-state updates depend only on the local GNN embedding $\bm z_i(t)$ and the local hidden state $\bm h_i(t)$. 

The remaining architectures incorporate the spatially aggregated hidden state $\overline{\bm h}_i(t)$ into different components of the recurrent unit. In model $C_3$, the aggregated hidden state enters the update and reset gates but is excluded from the candidate hidden-state update. Conversely, model $C_4$ incorporates $\overline{\bm h}_i(t)$ only into the candidate hidden-state update, while retaining conventional local gates. Finally, $C_5$ corresponds to the full GRNN architecture introduced in Eqs.~\eqref{eq:recurrent_gates} and~\eqref{eq:hidden_update}, in which the aggregated hidden state contributes to both the recurrent gates and the candidate hidden-state update.

\begin{table}
    \centering
    \caption{
        Comparison of the candidate architectures considered in
        the ablation study. All models use the same GNN backbone
        with feature dimensions $1\rightarrow128\rightarrow96
        \rightarrow64$. Recurrent models employ a hidden-state
        dimension of 96. The RMSE quantifies the discrepancy
        between the predicted temporal autocorrelation function
        and the KPM reference.
    }
    \label{tab:RMSE}
    \begin{ruledtabular}
    \begin{tabular}{lcccc}
        Model &
        GRU &
        $\overline{\bm h}_i$ in gates &
        $\overline{\bm h}_i$ in candidate &
        RMSE \\
        \hline
        $C_1$ & No  & --  & --  & 0.032842 \\
        $C_2$ & Yes & No  & No  & 0.018433 \\
        $C_3$ & Yes & Yes & No  & 0.006224 \\
        $C_4$ & Yes & No  & Yes & 0.017147 \\
        $C_5$ & Yes & Yes & Yes & \textbf{0.006148} \\
    \end{tabular}
    \end{ruledtabular}
\end{table}

The resulting temporal autocorrelation functions are shown in Fig.~\ref{fig:Abelation}, and their root-mean-square errors (RMSEs) relative to the KPM reference are summarized in Table~\ref{tab:RMSE}. The memoryless GNN ($C_1$) exhibits the largest discrepancy, with an RMSE of $3.28\times10^{-2}$. Introducing local recurrent memory ($C_2$) reduces this error to $1.84\times10^{-2}$, demonstrating the advantage of incorporating temporal information into the model. 

A more substantial improvement is observed for $C_3$, which incorporates spatially aggregated hidden states into the recurrent gates and achieves an RMSE of $6.22\times10^{-3}$. In contrast, incorporating the same spatial information only into the candidate hidden-state update ($C_4$) yields an RMSE of $1.71\times10^{-2}$, representing only a modest improvement over the conventional GRU. These results indicate that spatial coupling in the update and reset gates plays a particularly important role in reproducing the temporal correlations of the microscopic dynamics. 

The full GRNN architecture ($C_5$) achieves the lowest RMSE, $6.15\times10^{-3}$, although its improvement over $C_3$ is relatively small. This observation suggests that the spatially coupled recurrent gates are primarily responsible for the improved accuracy, while the additional coupling in the candidate hidden-state update provides a smaller incremental benefit. Overall, the ablation study supports the use of spatially coupled recurrent memory for learning the effective dynamics of the collective order paramete


\twocolumngrid

\bibliography{ref}

@article{yu2019,
  author  = {Yu, Yong and Si, Xiaosheng and Hu, Changhua and Zhang, Jianxun},
  title   = {A Review of Recurrent Neural Networks: {LSTM} Cells and Network Architectures},
  journal = {Neural Computation},
  volume  = {31},
  number  = {7},
  pages   = {1235--1270},
  year    = {2019},
  doi     = {10.1162/neco_a_01199},
}

@article{sherstinsky2020,
  author  = {Sherstinsky, Alex},
  title   = {Fundamentals of Recurrent Neural Network ({RNN}) and Long Short-Term Memory ({LSTM}) network},
  journal = {Physica D: Nonlinear Phenomena},
  volume  = {404},
  pages   = {132306},
  year    = {2020},
  doi     = {10.1016/j.physd.2019.132306},
}

@article{elman1990,
  author  = {Elman, Jeffrey L.},
  title   = {Finding Structure in Time},
  journal = {Cognitive Science},
  volume  = {14},
  number  = {2},
  pages   = {179--211},
  year    = {1990},
  doi     = {10.1207/s15516709cog1402_1},
}

@article{hochreiter1997,
  author  = {Hochreiter, Sepp and Schmidhuber, J{\"u}rgen},
  title   = {Long Short-Term Memory},
  journal = {Neural Computation},
  volume  = {9},
  number  = {8},
  pages   = {1735--1780},
  year    = {1997},
  doi     = {10.1162/neco.1997.9.8.1735},
}

@inproceedings{cho2014,
  author    = {Cho, Kyunghyun and van Merri{\"e}nboer, Bart and Gulcehre, Caglar and Bahdanau, Dzmitry and Bougares, Fethi and Schwenk, Holger and Bengio, Yoshua},
  title     = {Learning Phrase Representations using {RNN} Encoder--Decoder for Statistical Machine Translation},
  booktitle = {Proceedings of the 2014 Conference on Empirical Methods in Natural Language Processing ({EMNLP})},
  pages     = {1724--1734},
  year      = {2014},
  doi       = {10.3115/v1/D14-1179},
}

@misc{chung2014,
  author        = {Chung, Junyoung and Gulcehre, Caglar and Cho, Kyunghyun and Bengio, Yoshua},
  title         = {Empirical Evaluation of Gated Recurrent Neural Networks on Sequence Modeling},
  year          = {2014},
  eprint        = {1412.3555},
  archivePrefix = {arXiv},
  primaryClass  = {cs.NE},
  url           = {https://arxiv.org/abs/1412.3555},
}

@article{jaeger2004,
  author  = {Jaeger, Herbert and Haas, Harald},
  title   = {Harnessing Nonlinearity: Predicting Chaotic Systems and Saving Energy in Wireless Communication},
  journal = {Science},
  volume  = {304},
  number  = {5667},
  pages   = {78--80},
  year    = {2004},
  doi     = {10.1126/science.1091277},
}

@article{pathak2018,
  author  = {Pathak, Jaideep and Hunt, Brian and Girvan, Michelle and Lu, Zhixin and Ott, Edward},
  title   = {Model-Free Prediction of Large Spatiotemporally Chaotic Systems from Data: A Reservoir Computing Approach},
  journal = {Physical Review Letters},
  volume  = {120},
  pages   = {024102},
  year    = {2018},
  doi     = {10.1103/PhysRevLett.120.024102},
}

@article{chattopadhyay2020,
  author  = {Chattopadhyay, Ashesh and Hassanzadeh, Pedram and Subramanian, Devika},
  title   = {Data-driven predictions of a multiscale {Lorenz} 96 chaotic system using machine-learning methods: reservoir computing, artificial neural network, and long short-term memory network},
  journal = {Nonlinear Processes in Geophysics},
  volume  = {27},
  pages   = {373--389},
  year    = {2020},
  doi     = {10.5194/npg-27-373-2020},
}

@inproceedings{li2016,
  author    = {Li, Yujia and Tarlow, Daniel and Brockschmidt, Marc and Zemel, Richard},
  title     = {Gated Graph Sequence Neural Networks},
  booktitle = {International Conference on Learning Representations ({ICLR})},
  year      = {2016},
  eprint    = {1511.05493},
  archivePrefix = {arXiv},
  url       = {https://arxiv.org/abs/1511.05493}
}

@inproceedings{li2018,
  author    = {Li, Yaguang and Yu, Rose and Shahabi, Cyrus and Liu, Yan},
  title     = {Diffusion Convolutional Recurrent Neural Network: Data-Driven Traffic Forecasting},
  booktitle = {International Conference on Learning Representations ({ICLR})},
  year      = {2018},
  eprint    = {1707.01926},
  archivePrefix = {arXiv},
  url       = {https://arxiv.org/abs/1707.01926}
}

@article{scarselli2009,
  author  = {Scarselli, Franco and Gori, Marco and Tsoi, Ah Chung and Hagenbuchner, Markus and Monfardini, Gabriele},
  title   = {The Graph Neural Network Model},
  journal = {IEEE Transactions on Neural Networks},
  volume  = {20},
  number  = {1},
  pages   = {61--80},
  year    = {2009},
  doi     = {10.1109/TNN.2008.2005605},
}

@inproceedings{kipf2017,
  author    = {Kipf, Thomas N. and Welling, Max},
  title     = {Semi-Supervised Classification with Graph Convolutional Networks},
  booktitle = {International Conference on Learning Representations ({ICLR})},
  year      = {2017},
  eprint    = {1609.02907},
  archivePrefix = {arXiv},
  url       = {https://arxiv.org/abs/1609.02907}
}

@inproceedings{gilmer2017,
  author    = {Gilmer, Justin and Schoenholz, Samuel S. and Riley, Patrick F. and Vinyals, Oriol and Dahl, George E.},
  title     = {Neural Message Passing for Quantum Chemistry},
  booktitle = {Proceedings of the 34th International Conference on Machine Learning},
  series    = {Proceedings of Machine Learning Research},
  volume    = {70},
  pages     = {1263--1272},
  year      = {2017},
  publisher = {PMLR},
  url       = {https://proceedings.mlr.press/v70/gilmer17a.html}
}

@inproceedings{sanchezgonzalez2020,
  author    = {Sanchez-Gonzalez, Alvaro and Godwin, Jonathan and Pfaff, Tobias and Ying, Rex and Leskovec, Jure and Battaglia, Peter},
  title     = {Learning to Simulate Complex Physics with Graph Networks},
  booktitle = {Proceedings of the 37th International Conference on Machine Learning},
  series    = {Proceedings of Machine Learning Research},
  volume    = {119},
  pages     = {8459--8468},
  year      = {2020},
  publisher = {PMLR},
  url       = {https://proceedings.mlr.press/v119/sanchez-gonzalez20a.html}
}

@article{holstein1959,
  author  = {Holstein, Theodore},
  title   = {Studies of Polaron Motion: Part {I}. The Molecular-Crystal Model},
  journal = {Annals of Physics},
  volume  = {8},
  number  = {3},
  pages   = {325--342},
  year    = {1959},
  doi     = {10.1016/0003-4916(59)90002-8},
}

@article{scalettar1989,
  author  = {Scalettar, R. T. and Bickers, N. E. and Scalapino, D. J.},
  title   = {Competition of Pairing and Peierls--Charge-Density-Wave Correlations in a Two-Dimensional Electron-Phonon Model},
  journal = {Physical Review B},
  volume  = {40},
  number  = {1},
  pages   = {197--200},
  year    = {1989},
  doi     = {10.1103/PhysRevB.40.197},
  url     = {https://doi.org/10.1103/PhysRevB.40.197},
}

@article{weber2018,
  author  = {Weber, Manuel and Hohenadler, Martin},
  title   = {Two-Dimensional {Holstein--Hubbard} Model: Critical Temperature, {Ising} Universality, and Bipolaron Liquid},
  journal = {Physical Review B},
  volume  = {98},
  number  = {8},
  pages   = {085405},
  year    = {2018},
  doi     = {10.1103/PhysRevB.98.085405},
  url     = {https://doi.org/10.1103/PhysRevB.98.085405},
}

@article{noack1991,
  author  = {Noack, R. M. and Scalapino, D. J. and Scalettar, R. T.},
  title   = {Charge-Density-Wave and Pairing Susceptibilities in a Two-Dimensional Electron-Phonon Model},
  journal = {Physical Review Letters},
  volume  = {66},
  pages   = {778--781},
  year    = {1991},
  doi     = {10.1103/PhysRevLett.66.778},
}

@article{hohenadler2019,
  author  = {Hohenadler, Martin and Batrouni, G. G.},
  title   = {Dominant Charge Density Wave Correlations in the {Holstein} Model on the Half-Filled Square Lattice},
  journal = {Physical Review B},
  volume  = {100},
  pages   = {165114},
  year    = {2019},
  doi     = {10.1103/PhysRevB.100.165114},
}

@article{barros2013,
  author  = {Barros, Kipton and Kato, Yasuyuki},
  title   = {Efficient {Langevin} Simulation of Coupled Classical Fields and Fermions},
  journal = {Physical Review B},
  volume  = {88},
  pages   = {235101},
  year    = {2013},
  doi     = {10.1103/PhysRevB.88.235101},
}

@article{bray1994,
  author  = {Bray, A. J.},
  title   = {Theory of Phase-Ordering Kinetics},
  journal = {Advances in Physics},
  volume  = {43},
  number  = {3},
  pages   = {357--459},
  year    = {1994},
  doi     = {10.1080/00018739400101505},
}

@article{hohenberg1977,
  author  = {Hohenberg, P. C. and Halperin, B. I.},
  title   = {Theory of Dynamic Critical Phenomena},
  journal = {Reviews of Modern Physics},
  volume  = {49},
  pages   = {435--479},
  year    = {1977},
  doi     = {10.1103/RevModPhys.49.435},
}

@article{cahn1958,
  author  = {Cahn, John W. and Hilliard, John E.},
  title   = {Free Energy of a Nonuniform System. {I}. Interfacial Free Energy},
  journal = {The Journal of Chemical Physics},
  volume  = {28},
  number  = {2},
  pages   = {258--267},
  year    = {1958},
  doi     = {10.1063/1.1744102},
}

@article{allen1979,
  author  = {Allen, Samuel M. and Cahn, John W.},
  title   = {A Microscopic Theory for Antiphase Boundary Motion and Its Application to Antiphase Domain Coarsening},
  journal = {Acta Metallurgica},
  volume  = {27},
  number  = {6},
  pages   = {1085--1095},
  year    = {1979},
  doi     = {10.1016/0001-6160(79)90196-2},
}

@article{chen2002,
  author  = {Chen, Long-Qing},
  title   = {Phase-Field Models for Microstructure Evolution},
  journal = {Annual Review of Materials Research},
  volume  = {32},
  pages   = {113--140},
  year    = {2002},
  doi     = {10.1146/annurev.matsci.32.112001.132041},
}

@article{moreo1999,
  author  = {Moreo, Adriana and Yunoki, Seiji and Dagotto, Elbio},
  title   = {Phase Separation Scenario for Manganese Oxides and Related Materials},
  journal = {Science},
  volume  = {283},
  number  = {5410},
  pages   = {2034--2040},
  year    = {1999},
  doi     = {10.1126/science.283.5410.2034},
}

@article{moreo2000,
  author  = {Moreo, Adriana and Mayr, Matthias and Feiguin, Adrian and Yunoki, Seiji and Dagotto, Elbio},
  title   = {Giant Cluster Coexistence in Doped Manganites and Other Compounds},
  journal = {Physical Review Letters},
  volume  = {84},
  number  = {24},
  pages   = {5568--5571},
  year    = {2000},
  doi     = {10.1103/PhysRevLett.84.5568},
}

@article{white1998,
  author  = {White, Steven R. and Scalapino, D. J.},
  title   = {Density Matrix Renormalization Group Study of the Striped Phase in the {2D} {$t$}-{$J$} Model},
  journal = {Physical Review Letters},
  volume  = {80},
  number  = {6},
  pages   = {1272--1275},
  year    = {1998},
  doi     = {10.1103/PhysRevLett.80.1272},
}

@article{zheng2017,
  author  = {Zheng, Bo-Xiao and Chung, Chia-Min and Corboz, Philippe and Ehlers, Georg and Qin, Ming-Pu and Noack, Reinhard M. and Shi, Hao and White, Steven R. and Zhang, Shiwei and Chan, Garnet Kin-Lic},
  title   = {Stripe Order in the Underdoped Region of the Two-Dimensional {Hubbard} Model},
  journal = {Science},
  volume  = {358},
  number  = {6367},
  pages   = {1155--1160},
  year    = {2017},
  doi     = {10.1126/science.aam7127},
}

@article{huang2017,
  author  = {Huang, Edwin W. and Mendl, Christian B. and Liu, Shenxiu and Johnston, Steve and Jiang, Hong-Chen and Moritz, Brian and Devereaux, Thomas P.},
  title   = {Numerical Evidence of Fluctuating Stripes in the Normal State of High-{$T_c$} Cuprate Superconductors},
  journal = {Science},
  volume  = {358},
  number  = {6367},
  pages   = {1161--1164},
  year    = {2017},
  doi     = {10.1126/science.aak9546},
}

@article{chern2018,
  author  = {Chern, Gia-Wei and Barros, Kipton and Wang, Zhentao and Suwa, Hidemaro and Batista, Cristian D.},
  title   = {Semiclassical Dynamics of Spin Density Waves},
  journal = {Physical Review B},
  volume  = {97},
  number  = {3},
  pages   = {035120},
  year    = {2018},
  doi     = {10.1103/PhysRevB.97.035120},
}

@article{suwa2019,
  author  = {Suwa, Hidemaro and Smith, Justin S. and Lubbers, Nicholas and Batista, Cristian D. and Chern, Gia-Wei and Barros, Kipton},
  title   = {Machine Learning for Molecular Dynamics with Strongly Correlated Electrons},
  journal = {Physical Review B},
  volume  = {99},
  pages   = {161107},
  year    = {2019},
  doi     = {10.1103/PhysRevB.99.161107},
}

@article{zhang2021,
  author  = {Zhang, Puhan and Chern, Gia-Wei},
  title   = {Arrested Phase Separation in Double-Exchange Models: Large-Scale Simulation Enabled by Machine Learning},
  journal = {Physical Review Letters},
  volume  = {127},
  pages   = {146401},
  year    = {2021},
  doi     = {10.1103/PhysRevLett.127.146401},
}

@article{cheng2023a,
  author  = {Cheng, Chen and Zhang, Sheng and Chern, Gia-Wei},
  title   = {Machine Learning for Phase Ordering Dynamics of Charge Density Waves},
  journal = {Physical Review B},
  volume  = {108},
  pages   = {014301},
  year    = {2023},
  doi     = {10.1103/PhysRevB.108.014301},
}

@article{dinh2025,
  author  = {Dinh, Clement and Fan, Yunhao and Chern, Gia-Wei},
  title   = {Echo State Network for Coarsening Dynamics of Charge Density Waves},
  journal = {Physical Review E},
  volume  = {112},
  pages   = {065302},
  year    = {2025},
  doi     = {10.1103/w66k-fb13},
}

@article{fan2026,
  author  = {Fan, Yunhao and Chern, Gia-Wei},
  title   = {Graph Neural Network Force Fields for Adiabatic Dynamics of Lattice Hamiltonians},
  journal = {APL Machine Learning},
  volume  = {4},
  pages   = {036107},
  year    = {2026},
  doi     = {10.1063/5.0334142},
}

@article{bakshi2026,
  author  = {Bakshi, Sankha Subhra and Fan, Yunhao and Chern, Gia-Wei},
  title   = {Machine Learning Modeling of Charge-Density-Wave Recovery after Laser Melting},
  journal = {Physical Review B},
  volume  = {114},
  pages   = {165110},
  year    = {2026},
  doi     = {10.1103/klpt-f58d},
}

@article{zwanzig1961,
  author  = {Zwanzig, Robert},
  title   = {Memory Effects in Irreversible Thermodynamics},
  journal = {Physical Review},
  volume  = {124},
  pages   = {983--992},
  year    = {1961},
  doi     = {10.1103/PhysRev.124.983},
}

@article{mori1965,
  author  = {Mori, Hazime},
  title   = {Transport, Collective Motion, and Brownian Motion},
  journal = {Progress of Theoretical Physics},
  volume  = {33},
  number  = {3},
  pages   = {423--455},
  year    = {1965},
  doi     = {10.1143/PTP.33.423},
}

@article{chorin2000,
  author  = {Chorin, Alexandre J. and Hald, Ole H. and Kupferman, Raz},
  title   = {Optimal Prediction and the {Mori--Zwanzig} Representation of Irreversible Processes},
  journal = {Proceedings of the National Academy of Sciences USA},
  volume  = {97},
  number  = {7},
  pages   = {2968--2973},
  year    = {2000},
  doi     = {10.1073/pnas.97.7.2968},
}

@incollection{takens1981,
  author    = {Takens, Floris},
  title     = {Detecting Strange Attractors in Turbulence},
  booktitle = {Dynamical Systems and Turbulence, Warwick 1980},
  editor    = {Rand, David and Young, Lai-Sang},
  series    = {Lecture Notes in Mathematics},
  volume    = {898},
  pages     = {366--381},
  publisher = {Springer},
  address   = {Berlin, Heidelberg},
  year      = {1981},
  doi       = {10.1007/BFb0091924},
}

@article{herrera2024,
  author  = {Herrera Rodr{\'i}guez, Luis E. and Kananenka, Alexei A.},
  title   = {A Short Trajectory Is All You Need: A Transformer-Based Model for Long-Time Dissipative Quantum Dynamics},
  journal = {The Journal of Chemical Physics},
  volume  = {161},
  number  = {17},
  pages   = {171101},
  year    = {2024},
  doi     = {10.1063/5.0232871},
}

@misc{jang2026,
  author        = {Jang, Ho and Chern, Gia-Wei},
  title         = {Transformer Learning of Chaotic Collective Dynamics in Many-Body Systems},
  year          = {2026},
  eprint        = {2601.19080},
  archivePrefix = {arXiv},
  primaryClass  = {physics.comp-ph},
  doi           = {10.48550/arXiv.2601.19080},
}

@article{unke2021,
  author  = {Unke, Oliver T. and Chmiela, Stefan and Sauceda, Huziel E. and Gastegger, Michael and Poltavsky, Igor and Sch{\"u}tt, Kristof T. and Tkatchenko, Alexandre and M{\"u}ller, Klaus-Robert},
  title   = {Machine Learning Force Fields},
  journal = {Chemical Reviews},
  volume  = {121},
  number  = {16},
  pages   = {10142--10186},
  year    = {2021},
  doi     = {10.1021/acs.chemrev.0c01111},
}

@article{behler2007,
  author  = {Behler, J{\"o}rg and Parrinello, Michele},
  title   = {Generalized Neural-Network Representation of High-Dimensional Potential-Energy Surfaces},
  journal = {Physical Review Letters},
  volume  = {98},
  pages   = {146401},
  year    = {2007},
  doi     = {10.1103/PhysRevLett.98.146401},
}

@article{bartok2010,
  author  = {Bart{\'o}k, Albert P. and Payne, Mike C. and Kondor, Risi and Cs{\'a}nyi, G{\'a}bor},
  title   = {Gaussian Approximation Potentials: The Accuracy of Quantum Mechanics, without the Electrons},
  journal = {Physical Review Letters},
  volume  = {104},
  pages   = {136403},
  year    = {2010},
  doi     = {10.1103/PhysRevLett.104.136403},
}

@article{zhang2018,
  author  = {Zhang, Linfeng and Han, Jiequn and Wang, Han and Car, Roberto and E, Weinan},
  title   = {Deep Potential Molecular Dynamics: A Scalable Model with the Accuracy of Quantum Mechanics},
  journal = {Physical Review Letters},
  volume  = {120},
  pages   = {143001},
  year    = {2018},
  doi     = {10.1103/PhysRevLett.120.143001},
}

@article{batzner2022,
  author  = {Batzner, Simon and Musaelian, Albert and Sun, Lixin and Geiger, Mario and Mailoa, Jonathan P. and Kornbluth, Mordechai and Molinari, Nicola and Smidt, Tess E. and Kozinsky, Boris},
  title   = {{E(3)}-Equivariant Graph Neural Networks for Data-Efficient and Accurate Interatomic Potentials},
  journal = {Nature Communications},
  volume  = {13},
  pages   = {2453},
  year    = {2022},
  doi     = {10.1038/s41467-022-29939-5},
}

@article{zhang2023,
  author  = {Zhang, Puhan and Chern, Gia-Wei},
  title   = {Machine Learning Nonequilibrium Electron Forces for Spin Dynamics of Itinerant Magnets},
  journal = {npj Computational Materials},
  volume  = {9},
  pages   = {32},
  year    = {2023},
  doi     = {10.1038/s41524-023-00990-0},
}

@article{cheng2023b,
  author  = {Cheng, Xinlun and Zhang, Sheng and Nguyen, Phong C. H. and Azarfar, Shahab and Chern, Gia-Wei and Baek, Stephen S.},
  title   = {Convolutional Neural Networks for Large-Scale Dynamical Modeling of Itinerant Magnets},
  journal = {Physical Review Research},
  volume  = {5},
  pages   = {033188},
  year    = {2023},
  doi     = {10.1103/PhysRevResearch.5.033188},
}

@article{ghosh2024,
  author  = {Ghosh, Supriyo and Zhang, Sheng and Cheng, Chen and Chern, Gia-Wei},
  title   = {Kinetics of orbital ordering in cooperative {Jahn-Teller} models: Machine-learning enabled large-scale simulations},
  journal = {Physical Review Materials},
  volume  = {8},
  pages   = {123602},
  year    = {2024},
  doi     = {10.1103/PhysRevMaterials.8.123602},
}

@article{vlachas2018,
  author  = {Vlachas, Pantelis R. and Byeon, Wonmin and Wan, Zhong Y. and Sapsis, Themistoklis P. and Koumoutsakos, Petros},
  title   = {Data-Driven Forecasting of High-Dimensional Chaotic Systems with Long Short-Term Memory Networks},
  journal = {Proceedings of the Royal Society A},
  volume  = {474},
  number  = {2213},
  pages   = {20170844},
  year    = {2018},
  doi     = {10.1098/rspa.2017.0844},
}

@article{tsai2020,
  author  = {Tsai, Sun-Ting and Kuo, En-Jui and Tiwary, Pratyush},
  title   = {Learning Molecular Dynamics with Simple Language Model Built upon Long Short-Term Memory Neural Network},
  journal = {Nature Communications},
  volume  = {11},
  pages   = {5115},
  year    = {2020},
  doi     = {10.1038/s41467-020-18959-8},
}

@article{flurin2020,
  author  = {Flurin, Emmanuel and Martin, Leigh S. and Hacohen-Gourgy, Shay and Siddiqi, Irfan},
  title   = {Using a Recurrent Neural Network to Reconstruct Quantum Dynamics of a Superconducting Qubit from Physical Observations},
  journal = {Physical Review X},
  volume  = {10},
  pages   = {011006},
  year    = {2020},
  doi     = {10.1103/PhysRevX.10.011006},
}

@article{lin2021,
  author  = {Lin, Kunni and Peng, Jiawei and Gu, Feng Long and Lan, Zhenggang},
  title   = {Simulation of Open Quantum Dynamics with Bootstrap-Based Long Short-Term Memory Recurrent Neural Network},
  journal = {The Journal of Physical Chemistry Letters},
  volume  = {12},
  number  = {41},
  pages   = {10225--10234},
  year    = {2021},
  doi     = {10.1021/acs.jpclett.1c02672},
}

@article{ning2026,
  author  = {Ning, Alex and Yang, Lingyu and Chern, Gia-Wei},
  title   = {Recurrent Convolutional Neural Networks for Modeling Nonadiabatic Dynamics of Quantum-Classical Systems},
  journal = {Physical Review E},
  volume  = {113},
  pages   = {015307},
  year    = {2026},
  doi     = {10.1103/7dr5-68ky},
}

@article{koppe2019,
  author  = {Koppe, Georgia and Toutounji, Hazem and Kirsch, Peter and Lis, Stefanie and Durstewitz, Daniel},
  title   = {Identifying nonlinear dynamical systems via generative recurrent neural networks with applications to {fMRI}},
  journal = {PLOS Computational Biology},
  volume  = {15},
  number  = {8},
  pages   = {e1007263},
  year    = {2019},
  doi     = {10.1371/journal.pcbi.1007263},
}

@article{vlachas2022,
  author  = {Vlachas, Pantelis R. and Arampatzis, Georgios and Uhler, Caroline and Koumoutsakos, Petros},
  title   = {Multiscale simulations of complex systems by learning their effective dynamics},
  journal = {Nature Machine Intelligence},
  volume  = {4},
  number  = {4},
  pages   = {359--366},
  year    = {2022},
  doi     = {10.1038/s42256-022-00464-w},
}

@article{banchi2018,
  author  = {Banchi, Leonardo and Grant, Edward and Rocchetto, Andrea and Severini, Simone},
  title   = {Modelling non-{Markovian} quantum processes with recurrent neural networks},
  journal = {New Journal of Physics},
  volume  = {20},
  number  = {12},
  pages   = {123030},
  year    = {2018},
  doi     = {10.1088/1367-2630/aaf749},
}

@article{durstewitz2023,
  author  = {Durstewitz, Daniel and Koppe, Georgia and Thurm, Max Ingo},
  title   = {Reconstructing computational system dynamics from neural data with recurrent neural networks},
  journal = {Nature Reviews Neuroscience},
  volume  = {24},
  number  = {11},
  pages   = {693--710},
  year    = {2023},
  doi     = {10.1038/s41583-023-00740-7},
}

@article{zhang2022,
  author  = {Zhang, Sheng and Zhang, Puhan and Chern, Gia-Wei},
  title   = {Anomalous Phase Separation in a Correlated Electron System: Machine-Learning-Enabled Large-Scale Kinetic {Monte Carlo} Simulations},
  journal = {Proceedings of the National Academy of Sciences},
  volume  = {119},
  number  = {18},
  pages   = {e2119957119},
  year    = {2022},
  doi     = {10.1073/pnas.2119957119},
}

@article{fan2024,
  author  = {Fan, Yunhao and Zhang, Sheng and Chern, Gia-Wei},
  title   = {Coarsening of Chiral Domains in Itinerant Electron Magnets: A Machine Learning Force-Field Approach},
  journal = {Physical Review B},
  volume  = {110},
  pages   = {245105},
  year    = {2024},
  doi     = {10.1103/PhysRevB.110.245105},
}

@article{tyberg2025,
  author  = {Tyberg, Alexa and Fan, Yunhao and Chern, Gia-Wei},
  title   = {Machine Learning Force Field Model for Kinetic {Monte Carlo} Simulations of Itinerant {Ising} Magnets},
  journal = {Physical Review B},
  volume  = {111},
  pages   = {235132},
  year    = {2025},
  doi     = {10.1103/d3zm-pbr1},
}

@article{chern2026aip,
  author  = {Chern, Gia-Wei and Fan, Yunhao and Zhang, Sheng and Zhang, Puhan},
  title   = {Machine-Learning Force-Field Models for Dynamical Simulations of Metallic Magnets},
  journal = {AIP Advances},
  volume  = {16},
  number  = {2},
  pages   = {025031},
  year    = {2026},
  doi     = {10.1063/9.0001056},
}

@article{chern2026jmmm,
  author  = {Chern, Gia-Wei and Fan, Yunhao and Zhang, Sheng and Zhang, Puhan},
  title   = {Machine-Learning Modeling of Magnetization Dynamics in Quasi-Equilibrium and Driven Metallic Spin Systems},
  journal = {Journal of Magnetism and Magnetic Materials},
  volume  = {642},
  pages   = {173898},
  year    = {2026},
  doi     = {10.1016/j.jmmm.2026.173898},
}

@inproceedings{batatia2022,
  author    = {Batatia, Ilyes and Kov{\'a}cs, D{\'a}vid P{\'e}ter and Simm, Gregor N. C. and Ortner, Christoph and Cs{\'a}nyi, G{\'a}bor},
  title     = {{MACE}: Higher Order Equivariant Message Passing Neural Networks for Fast and Accurate Force Fields},
  booktitle = {Advances in Neural Information Processing Systems},
  volume    = {35},
  pages     = {11423--11436},
  year      = {2022},
  doi       = {10.52202/068431-0830},
  url       = {https://papers.neurips.cc/paper_files/paper/2022/hash/4a36c3c51af11ed9f34615b81edb5bbc-Abstract-Conference.html},
}

@misc{bora2026cooperative,
  author        = {Bora, Arunangshu and Bakshi, Sankha Subhra and Jang, Ho and Chern, Gia-Wei},
  title         = {Cooperative Domain-Wall Dynamics in a Two-Dimensional Quasiclassical {Holstein} Model},
  year          = {2026},
  eprint        = {2609.28436},
  archivePrefix = {arXiv},
  doi           = {10.48550/arXiv.2609.28436},
  url           = {https://arxiv.org/abs/2609.28436},
}

@article{jang2026kinks,
  author  = {Jang, Ho and Yang, Yang and Chern, Gia-Wei},
  title   = {Anomalous Coarsening and Nonlinear Diffusion of Kinks in a One-Dimensional Quasiclassical {Holstein} Model},
  journal = {Physical Review E},
  volume  = {114},
  number  = {1},
  pages   = {014225},
  year    = {2026},
  doi     = {10.1103/gc41-d1yq},
}

@article{weisse2006,
  author  = {Wei{\ss}e, Alexander and Wellein, Gerhard and Alvermann, Andreas and Fehske, Holger},
  title   = {The Kernel Polynomial Method},
  journal = {Reviews of Modern Physics},
  volume  = {78},
  number  = {1},
  pages   = {275--306},
  year    = {2006},
  doi     = {10.1103/RevModPhys.78.275},
}

\end{document}